# Balancing Act: The Cost of Wind Restrictions in Norway's Electricity Transition

Maximilian *Roithner**[1], Paola *Velasco Herrejon*[1], Koen *van Greevenbroek*[4], Aleksander *Grochowicz*[2,3], Oskar *Vågerö*[1], Tobias *Verheugen Hvidsten*[1], James *Price*[5], Marianne *Zeyringer*[1]

[1]Department of Technology Systems, University of Oslo

[2]Department of Mathematics, University of Oslo

[3]Department of Wind and Energy Systems, Technical University of Denmark

[4]Department of Computer Science, UiT The Arctic University of Norway

[5]UCL Energy Institute, University College London, London, United Kingdom

## Abstract

To meet its commitments under the Paris Agreement and reduce its dependency on energy imports, the pace, and scale of renewable energy deployment across Europe must increase dramatically over the next decade. Such a steep change in the net-zero transition will inevitably necessitate trade-offs with other societal priorities. Here we investigate a case study focused on the opposition towards onshore wind and the compromises that may need to be made to deliver its plans for deep electrification. Using an electricity system model, we explore the implications of key social and environmental dimensions shaping the future deployment of onshore wind on the costs and design of electricity systems for Norway in 2030. We find that under restrictions that allow for almost no additional onshore wind, demand can not be met and load has to be shed. Yet, when reducing the restrictions on onshore wind or allowing for in-country transmission expansion, feasible system designs at a small fraction of that cost can be found. To meet the net-zero targets, compromises will need to be made on either wind power deployment, transmission expansion, non-electrification of industry or demand reduction.

* corresponding author, maximilian.roithner@its.uio.no

# 1. Introduction

The decarbonization of power production is key to achieving the Paris Agreement goal of limiting global mean surface temperature rise to well below 2 °C, particularly so given the drive to electrify industry, transport, and heat. Variable renewable energy technologies (VREs) such as wind and solar photovoltaics (PV) have decreased rapidly in cost and matured into cost-effective decarbonization solutions (IPCC, 2023). However, the location of VREs affects the technical feasibility and their impact on the environment and the communities where they are located. Thus, socio-environmental constraints can have a large impact on the overall capacity potential which influences the technology choices, changing costs and political viability of reaching decarbonization goals. These constraints may also have an impact on ensuring a reliable electricity supply that meets current and future energy demand. The potential trade-offs and competing interests between technology, nature protection, social acceptance, energy prices and future demand needs to be evident in policies aimed at promoting VREs. There is an urgency for rapid action to close the emission gap requiring cuts of 42% by 2030 to get on track for 1.5 °C (United Nations Environment Programme et al., 2024) and e.g. the European Union is not on track to meet its 2030 targets (Climate Analytics & NewClimate Institute, 2024).

Norway comes with some of the best on- and offshore wind resources in Europe (Egging & Tomasgard, 2018; Karlstrøm & Ryghaug, 2014). If Norway aims to achieve net-zero emissions, this would lead to an increase in domestic electricity demand of up to 90 TWh from 127 TWh in 2023 (Statnett, 2023b) to electrify sectors including transport, manufacturing, and oil and gas extraction. However, wind energy development has been contested and licensing has been revoked due to opposition from nature conservation groups, recreational activities and local communities (Gulbrandsen et al., 2021; Karlstrøm & Ryghaug, 2014). While renewable energy development, nature protection and social support may all be regarded as critical, an energy transition that balances all three is proving challenging in meeting Norway's future energy demand and achieving its decarbonization goals. For instance, the construction of wind farms has been considered by the Sámi Council as threatening the sustainability of reindeer herding (Lawrence, 2014). In 2021, Norway's supreme court ruled that two wind farms built at Fosen in central Norway violated Sámi human rights under international conventions. The future of these wind farms is still unclear. This uncertainty sparked protests in February 2023, where Sámi activists blocked the entrance to Norway's energy ministry, demanding the cease of operations of the energy plants (Fouche & Klesty, 2023). Therefore, socio-environmental constraints can have a large impact on the overall electricity generation capacity potential, which will in turn affect optimal technology choices, system costs and the social feasibility of reaching the Paris Agreement. The Norwegian power system is facing an increase in electricity demand from the electrification of transport, heating and industry, while the traditional generation source (hydropower) is not able to meet all of this increase due to environmental limits. While there has been a strong increase in onshore wind energy capacity from 860 MW in 2014 to 5 GW in 2021, there have been only 15 MW added since then (Statistics Norway, 2024). This leads to concerns about a power deficit by 2030 (Statnett, 2023a) which we choose as the target year of the analysis.

Energy systems and electricity system modelling has been a key policy tool for studying how to meet future demand and decarbonization pathways (DeCarolis et al., 2017). They can provide knowledge-based and systematic methods and solutions to reach decisions about which technologies and areas to invest in. Nevertheless, present-day models mainly integrate techno-economic input parameters, whereas social factors and environmental constraints such as local acceptance of new installations, are largely neglected (Gambhir, 2019; Nikas et al., 2020; Pfenninger et al., 2014). Several studies (Höltinger et al., 2016; McKenna et al., 2014; Permien & Enevoldsen, 2019; Rinne et al., 2018) have acknowledged the importance of accounting for socio-environmental acceptance when modelling renewable energy potential and others call for integration of non techno-economic factors in energy system models (Hanna & Gross, 2021; Hirt et al., 2020; Süsser et al., 2022). Without considering social factors that shape the renewable energy deployment, energy system models can therefore produce decarbonization solutions that are neither publicly nor politically feasible (Trutnevyte, 2016), risking missing carbon targets.

In the last 10 years, modellers have incorporated socio-technical assumptions which have increased the complexity of models through the inclusion of social aspects (Krumm et al., 2022): Bolwig et al. (2020) use an energy systems model to assess the costs of social acceptance limiting the expansion of onshore wind and transmission capacities for the Nordic-Baltic region in 2030 and 2050. They do not perform a spatial analysis but develop four scenarios where transmission and/or onshore wind energy can be expanded. Price et al. (2018) assess how social and environmental restrictions on nuclear/renewables siting shape Great Britain's 2050 power system. Cheng et al. (2024) assess the case for Norwegian hydrogen exports, developing three scenarios including socio-environmental factors such as land-use and electricity prices for 2050. Inderberg et al. (2024) combines energy system optimization modelling with political feasibility of different transition pathways. They develop a scenario for Norway towards 2050 that is unrestrained by assumptions about policy, and based on that identify areas where political choices are key to model outcomes. Grimsrud et. al (2023) integrate monetized local disamenity and carbon sequestration costs and place constraints on areas of importance for wilderness and biodiversity for onshore wind deployment into a Norwegian energy system model for 2050. They only consider locations where concessions have been applied and limit expansion to an increase in onshore wind capacity of maximum 4 TWh annual production, adding to 15.5 TWh produced today.

Hirt et al. (2020) identified the need for integrative research to provide more practical outcomes to meet energy and climate targets. As energy policy and infrastructure decisions are made on a national level and 2030 is within today's politicians' timeline, this study provides practical, short-term policy-relevant insights on trade-offs and compromises. It is a socio-political decision to select more expensive technologies, sites, or mitigation options to minimize the socio-environmental impacts of VRE development. Yet, a spatially explicit capacity assessment under different socio-environmental scenarios combined with energy system modelling is missing for 2030 to allow for such discussion. Further, most modelling approaches do not capture the spatial detail of capacity (i.e. how much can be built in a region) as well as spatio-temporal production (i.e. how much can be produced hourly in that location) to account for the spatio-temporal variability of renewables.

Here, we close this gap by performing a nationally specific analysis: we first study the NVE (Norwegian Water Resources and Energy Directorate) framework, previous licences, literature, and newspaper articles to design three levels of socio-environmental acceptance (we call them "None", "Low" and "High") for onshore wind. Based on the developed levels, we then conduct a GIS analysis to determine the spatially dependent capacity potential per level of each studied dimension (Nature, Fauna, Sami, Neighbour). A description of the dimensions can be found in section 2.2.1. These dimensions contribute to building nationally specific socio-environmental scenarios that help decide which projects and locations can be considered, taking into account costs, future demand and decarbonization goals. We use these spatially explicit scenarios in an electricity system model and run it at a high spatial resolution (30 km) for variable renewable production to capture the spatio-temporal variability. We identify trade-offs and support policymakers with quantification on alternative compromises.

Due to its considerable VRE potential, socio-environmental opposition, and expected future growth in electricity demand, the case study therefore contributes to addressing the significant gap in identifying and analysing social and environmental variables. These variables can affect system design, costs, and therefore prices and exports. We do this by answering the following research questions:

- What technical, environmental and social factors can impact the land availability for wind energy in Norway?
- How do spatially dependent technical, environmental and social scenarios change Norway's cost-optimal wind energy capacity potential?
- How do these scenarios impact the cost, optimal design, unmet electricity demand and electricity imports of Norway's energy system in 2030?

To answer these research questions, section 2 describes the methodology including the model, data and scenarios, section 3 shows the results and finally, Section 4 discusses these findings and provides policy recommendations. While this paper draws on the Norwegian energy system, the globalized nature of both social and environmental restrictions and energy prices means that the research in Norway is likely to be relevant elsewhere.

## 2. Methodology

First, we describe the model, next the criteria for excluding areas from onshore wind instalment and distinguishing the different levels, and finally we show the area that remains available after applying those criteria.

All values with the unit Euro (€/EUR) refer to the value of that currency in the year 2023.

The input data, as well as results and the code used to generate them, can be found on Zenodo.

## 2.1. Model description

We separate between the model and its configuration in the first part and the generation of input data for the model in the second part of this section.

### 2.1.1. highRES electricity system model

We employ a modified version of the highRES electricity system model (Moore et al., 2018; Price et al., 2018, 2022, 2023; Price & Zeyringer, 2022; Zeyringer, Fais, et al., 2018; Zeyringer, Price, et al., 2018). highRES describes a linear optimization problem implemented in the *General algebraic modeling system* (GAMS) and is solved using the off-the-shelf mathematical program solver suite CPLEX for minimal total system cost, consisting of annualized investment and operational cost. A set of technical, economic, meteorological and land use constraints ensures operational feasibility under the given circumstances. Using perfect operational foresight with an hourly time resolution, we adapt it to represent the Norwegian power system on a NUTS level 3 (based on 2021), which corresponds to 11 administrative regions at that time. Supply and demand are balanced at the NUTS level through the transmission grid, but the model can deploy wind and solar capacity in the most optimal 30 km x 30 km grid cells (based on ERA5 data). We assume a fully decarbonized power grid (as is the case with domestic generation) by 2030 and therefore only include feasible zero-carbon technologies. Wind, Lithium battery storage and solar power, are technologies that can be expanded, whereas hydropower and pumped hydro storage are fixed to current capacities.

We place an upper limit on solar deployment due to low installation volumes and rates, compared to other technologies in the power system. The values can be found in table CAPLIM in the appendix. Furthermore, we assume that solar can not be built in areas that have protection level 1-IV as defined by the *International Union for Conservation of Nature and Natural Resources* (IUCN), slope greater than 15 degrees, elevation of more than 2000 metres above sea level. Also excluded are areas that belong to land cover categories, as supplied by the *Coordination of Information on the Environment* (CORINE) program, that are part of a list found in the appendix.

For offshore wind, we exclude 1a-IV of the World database of protected areas and areas deeper than 70 metres where bottom-mounted can be built, leading to an upper limit of 125 GW installable bottom-mounted offshore wind capacity. Norway today has no commercial offshore wind and since the Norwegian stated target for 2030 is 3 GW, in the waters before Rogaland, we limit the model to this value (Aasland, 2023; Buljan, 2022; Lange & Pochhammer, 2024). We do not include floating offshore wind because of the lead time of installations, which would make it difficult to finish large capacities until the target year of the analysis (2030).

In the partial greenfield optimization of Norway we include existing infrastructure such as transmission, power generation capacities (in particular 5 GW of onshore wind power and 30 GW of current hydropower) and pumped hydro storage (1.3 GW which will be operational in 2030). For existing infrastructure, only variable O&M costs are part of the cost minimization.

We assume a value of lost load of approximately €23 thousand/MWh based on Ovaere et al. (2016). The value of lost load tries to capture the economic impact of power deficits in the system. We assume electricity demand to increase from 140 TWh (in 2022) to 178 TWh in 2030 based on forecasts by the transmission system operator Statnett (2023a). A more detailed description of this can be found in the appendix in section 6.4. Given the limited time for grid expansion until 2030 both for in-country transmission and import capacity, and a desire to keep import volumes similar to current ones, we place an upper limit on the power capacity of both. A table with the limit values can be found in the appendix. A fixed price for imports of €34/megawatt hour(MWh) is assumed based on 2020 hourly electricity prices from the countries where there is an interconnection with Norway, weighted by the interconnector size.

More information on the modelling approach and assumptions can be found in section 6.2 of the appendix.

### 2.1.2. Capacity factor modelling based on weather data

We use the open-source tool atlite (Hofmann et al., 2021) to convert weather variables to power system variables (capacity factors, etc.), weighted by geographical availability (topography, land use, and the restricted land use scenarios described in Section 2.3.1).

To represent the influence of inter-annual weather variability on the system design (Grochowicz et al., 2023), we pick a challenging year (i.e. high total system costs) for the Norwegian electricity system. The year 2010 was characterized by low hydropower production and a very cold winter, leading to high electricity demand.

The complex topography of Norway means that the original ERA5 reanalysis weather data (Hersbach, H. et al., 2018) at 0.25 degree grid size cannot capture local variations in wind speeds, leading to a priori underestimation of wind power capacity factors in suitable locations. Therefore, we compute bias-correction functions for each 0.25 degree x 0.25 degree grid cell by comparing actual wind power production data and ERA5 wind speeds from 2019. The historical production data is sourced from the Norwegian Water Resources and Energy Directorate (NVE) (2023) for Norwegian wind parks constructed up to 2021; for wind parks constructed after 2019 the production data reported by NVE is based on regional climate modelling. For each wind park, we compute the year-round distributions of capacity factors based on ERA5 wind speeds at that wind park and those from historical production data, and find a bias-correction function $f\colon [0, 1] \to [0, 1]$ such that the distribution of bias-corrected ERA5 capacity factors matches the distribution of historical capacity factors for that wind park. For every grid cell, we then take as the bias-correction function the inverse distance weighted average of the bias-correctors for the 10 closest wind parks.

We exclude areas with known low wind speeds (less than 6.5 metres/second in 120m height), defined as hard exclusions "wind speed" NVE and therefore low wind power production potential (Norwegian Water Resources and Energy Directorate, 2019b).

## 2.2. Scenario design

We start by explaining the scenario dimensions and then detail how we combine the different levels of those dimensions into names for the resulting scenarios.

### 2.2.1. Land Exclusions dimensions

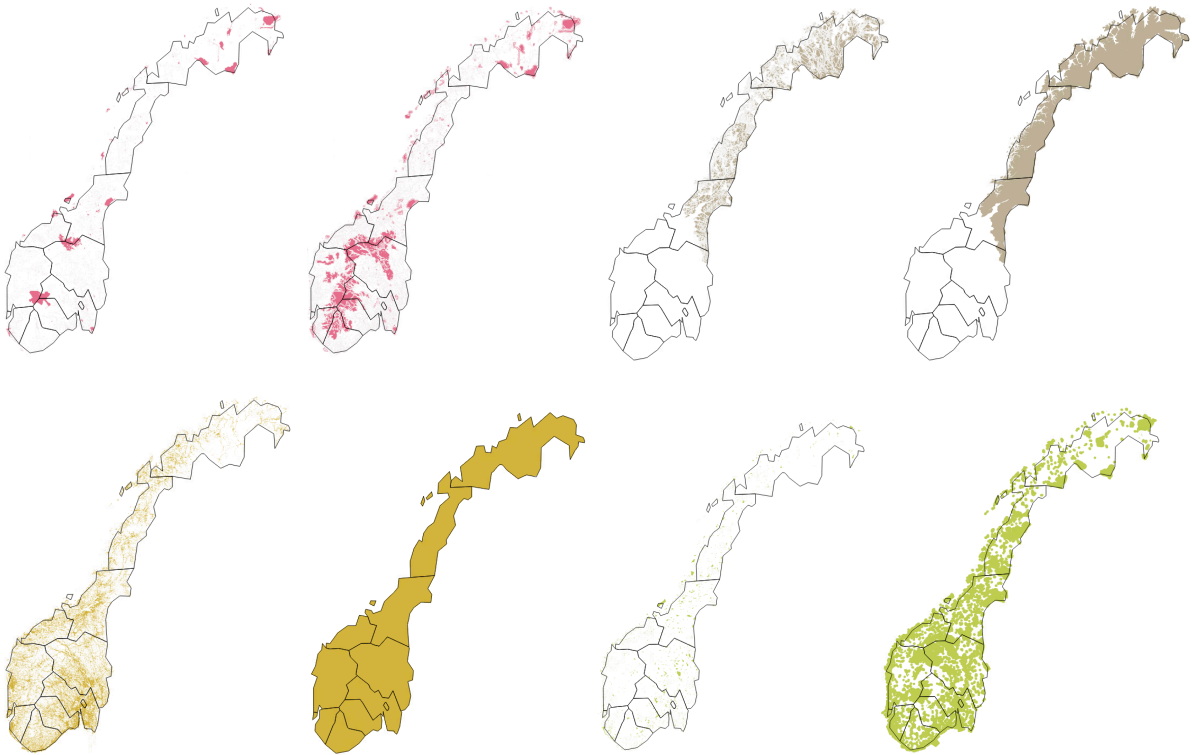

Figure EXCL: Plots of Low and High level (left to right) of exclusions for dimensions (from top left, one colour per dimension) Fauna, Sami, Neighbour and Nature.

We define three levels restricting the build-out of onshore wind in Norway until 2030 for four different restriction dimensions (technical, fauna, Sámi, neighbours). Level **None** considers only the technical constraints. Level **Low** adds environmental and social constraints to the existing technical factors, and level **High** considers even higher environmental and social constraints. Later, we create scenarios combining one expression for each dimension. Figure EXCL shows the land that is excluded from development for the **Low** and **High** level for each of the restriction dimensions.

The selection is informed by policy documents including NVE's onshore wind power framework (Norwegian Water Resources and Energy Directorate, 2019a), NVE's consequence report for offshore wind power (2013) and NVE's licence decision in onshore and offshore wind power cases. Furthermore, factors included in NVE licence decisions, and legal conventions that apply to the installation of renewable energy, such as the Bonn and Bern conventions.

In Tables 1 to 5 we describe the constraints considered for each of the levels (**None**, **Low**, **High**) for onshore wind power in Norway, as well as the reasoning behind the selection of each constraint. These include untouched connected nature areas, landscape, cultural heritage, outdoor life and tourism, nature types, fauna such as birds, bats, predators, wild reindeer and reindeer herding. Neighbour effects such as visibility and noise, and local and regional commerce development are also included.

### 2.3.1.1. Technical

| |
|---|
| Roads and Railway – 200m buffer |
| Airports 2 km buffer |
| Watercourses, water bodies and glaciers |
| Gradients steeper than 15 degrees, areas 2000 m or more above sea level |

**Table 1: Technical restrictions**

A buffer of 200m is applied to roads and railway in case a wind turbine falls or breaks (Enevoldsen & Permien, 2018). For airports, a 2 km buffer was added due to safety. Watercourses, water bodies and glaciers were excluded since these would require offshore wind turbines and foundations. Gradients steeper than 20 degrees were also excluded due to the difficulties of accessing these sites (Permien & Enevoldsen, 2019) and because due to technical and fluid-mechanical reasons they are less suitable for wind turbines (McKenna et al., 2014).

### 2.3.1.2. Nature

| **Low** | **High** |
|---|---|
| Strict nature reserve (IUCN category 1a) | Strict nature reserve (all IUCN categories + 5 km buffer) |
| Coastal heather | Coastal heather + 5 km buffer |

**Table 2:** Nature-related restrictions

As well as having an intrinsic value, natural reserves benefit various social and environmental factors such as nature types, fauna such as birds, bats, predators and non-domesticated reindeer, landscape, cultural heritage, outdoor life (Outdoor Recreation Act, 1957) and tourism. As well as being legally protected (Ot.Prp. Nr. 52 (2008-2009), 2009), conserving nature and the landscape for future generations is strongly rooted in Norwegian culture (Norwegian Environment Agency, 2019a) and politics (Ministry of Climate and Environment, 2015). Landscapes and nature types also define one's identity to a place and a cultural heritage (Norwegian Environment Agency, 2019b). The visual effects on the landscape and for cultural heritage, along with noise, have been identified as some of the most important disadvantages of wind farms in Norway (Norwegian Water Resources and Energy Directorate, 2012).

23 areas for the endangered nature type coastal heather are appointed for conservation and excluded according to the NVE's criteria (Norwegian Water Resources and Energy Directorate, 2019a)

### 2.3.1.3. Fauna

| Low | High |
|---|---|
| Important Bird Areas (IBA) as defined by Birdlife international. | 3 km buffer zone around IBA-areas |
| Four big predators (wolf, bear, lynx, and wolverine) and the mountain fox | Four big predators (wolf, bear, lynx, and wolverine) and the mountain fox<br><br>All areas of important biodiversity as defined by the Norwegian Environment Agency |
|  | All areas where wild reindeer live. All the 23 function areas for wild reindeer, including connected nature areas. |

**Table 3:** Fauna-related restrictions

Fauna, in the context of onshore wind power development, encompasses various species such as birds and other animals. Specifically, through the different levels, we have included birds, bats, wild reindeer and four major predators: wolves, bears, lynxes, and wolverines. Additionally, the fauna restriction includes mountain foxes, deer species like moose and deer, amphibians, and small rodents (Norwegian Water Resources and Energy Directorate, 2018).

There is little research indicating that wind power installations have negative effects on the bird species on the stock level (Norwegian Water Resources and Energy Directorate, 2018), which is why bird-related restrictions are not included in Scenario Two. Studies indicate that the amount of birds that collide with wind turbines is very low compared to other mortality factors created by humans (Norwegian Water Resources and Energy Directorate, 2018). Following the principle of precaution, the uncertainty about collision risk and repression results in restricting areas with a large bird population. Furthermore, IBA-areas are important bird areas which are not suitable for wind power (Rydell et al., 2012). The effects on birds on an individual level will not be taken into consideration by NVE and are therefore not restricted in our map. Due to the lack of concrete Norwegian recommendations for buffer zones around bird areas, we use the 3 km used by the Swedish Environmental Protection Agency (Rydell et al., 2012).

In protected national reserves, the presence of wild reindeer alone can be a sufficient reason to reject a concession application. As a result, areas with wild reindeer are excluded in Scenarios Two and Three. Non-Sámi domesticated reindeer herding is practised in parts of Innlandet county, with historical roots dating back to the 1700s. These reindeer herders operate within the framework of the Reindeer Herding Act (Reindeer Herding Act, 2021).

### 2.3.1.4. Sámi reindeer herding (domesticated reindeer)

| Low | High |
|---|---|
| Migratory zone (ensure the migration between different grazing land during the year cycle) | All areas used by reindeer (Migratory zones, all seasons grazing land, concession permit regions, Expropriation areas, enclosed grazing sites, and gathering sites). Reindeer herding demands big, connected areas, where other activities can disturb the reindeer and affect the herding. |
| Winter grazing land (Minimum grazing land areas) | |
| Core areas of the summer grazing land area | |
| Mating land / Autumn grazing land, unless during the reproduction time | |
| Calving area | |

**Table 4:** Reindeer-related restrictions

Migratory zones are protected according to the Reindeer Herding Act § 22. These areas ensure the migration between different grazing lands during the year cycle. NVE considers it unlikely to find acceptable alternative areas, so compensation could be difficult to establish (Norwegian Water Resources and Energy Directorate, 2018).

### 2.3.1.4. Neighbour effects (includes local and regional commerce development)

| Low | High |
|---|---|
| Noise (Buildings — a 400m buffer zone to keep noise levels under 50 dB). | Noise (Buildings — a buffer zone of 10 times the height of the wind turbine to keep noise levels under 40 dB). |
| Visibility (agglomerations): 1 km from cities and populated areas. | Visibility (agglomerations): 50 km from cities and populated areas. |
| Walking routes, ski runs, biking routes other routes — 20m buffer | |

**Table 5:** Neighbour-related restrictions

In 2014 a million Norwegians lived in buildings with noise levels over the limit value for traffic noise (Lden 55 dBA). Still, studies indicate that noise from wind turbines can be more troublesome than noise from traffic (Katinas et al., 2016). The recommended noise value by NVE and the Norwegian Environment Agency is 45 dBA (Ministry of Climate and Environment, 2021). According to the Norwegian Institute of Public Health, annoyance starts at levels over 40 dB (Norwegian Water Resources and Energy Directorate, 2018). To avoid this level, a buffer zone of 10 times the wind turbines' height was set (as is the case in countries such as Poland and Germany) (Norwegian Water Resources and Energy Directorate, 2018).

For visibility impacts, the Norwegian topography minimizes the consequences for neighbours, since flat terrain is rare. The restriction zone was set under 1 km, which considers at least three times the height of the wind turbine. Minimum standards in other countries are 500m in Ireland, ten times the wind turbines' height in Poland and Bavaria, Germany, and four times the height in Denmark (Norwegian Water Resources and Energy Directorate, 2018). The distance criteria should be determined in connection with the height of the wind turbine, this also ensures that the minimum distance is not too short for higher wind turbines (Norwegian Water Resources and Energy Directorate, 2018).

### 2.2.2. Scenario naming

The figures in this paper use a scenario naming based on the different varying levels for each dimension (i.e. a scenario consisting of a selection of None/Low/High values for four different dimensions). Their meaning can be decoded following the examples given in table SCENNAME. It shows the pattern, which is just one word if all dimensions (Nature, Fauna, Sámi, Neigh) are at the same level (**None** if all of them are None, **Low** if all of them are Low). The default assumption for all other scenarios except for those two is that all dimensions are at the "Low" level, and the display name contains the dimensions that are "High". To study the impacts of strict constraints on onshore wind, we ignore combinations of the levels "Low" and "None". This results in the following ten scenarios: None; Low; Sami; Fauna; Fauna,Sami; Nature; Nature,Sami; Nature,Fauna; Nature,Fauna,Sami; Nature,Fauna,Sami,Neigh;

| *Land Use* **Display Name** | *Nature* | *Fauna* | *Sámi* | *Neigh* |
|---|---|---|---|---|
| **None** | None | None | None | None |

| **Low** | Low | Low | Low | Low |
|---|---|---|---|---|
| **Nature** | High | Low | Low | Low |
| **Fauna** | Low | High | Low | Low |
| **Nature, Fauna** | High | High | Low | Low |
| **…** | … | … | … | … |

**Table SCENNAME:** Exemplary translation table between the scenario name and the scenario dimension level values

## 2.3. Area

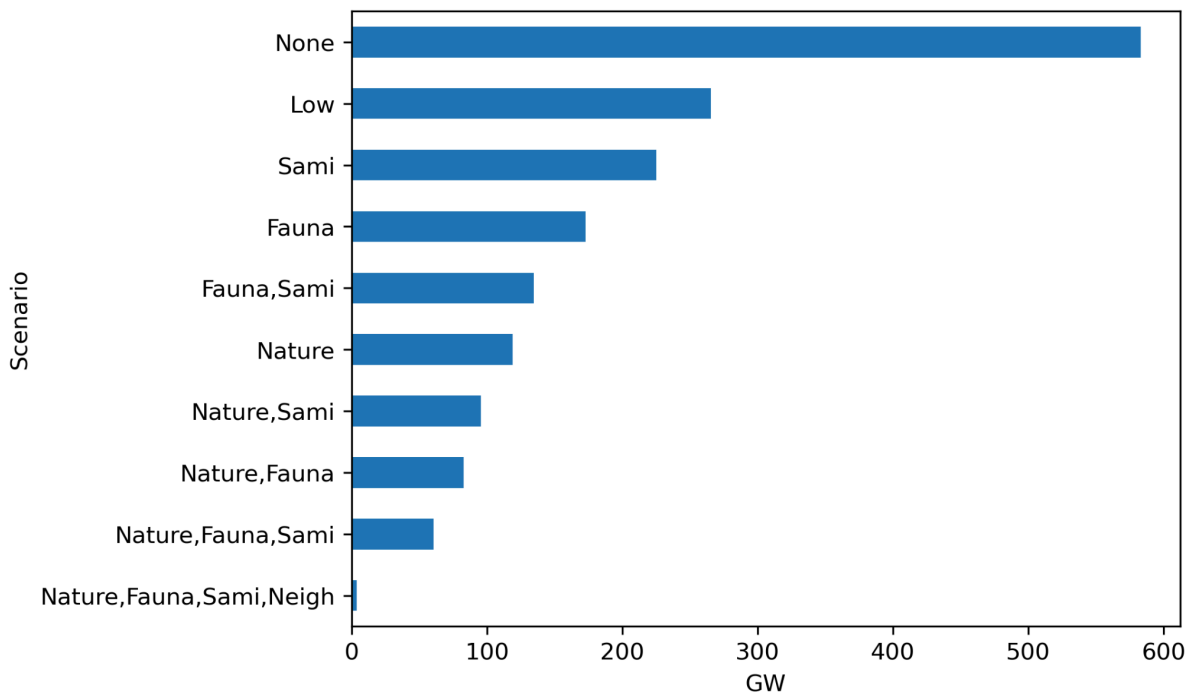

Figure AREASABS: Energy equivalent land area of Norway available for wind power, after land use restrictions have been applied in gigawatts (GW)

Figure AREASABS shows the available area for onshore wind in each of the land use restriction scenarios. The unit is gigawatts (GW) to make it easier to compare this figure with the results later. We assume that one can install 3 megawatt (MW)/km² (Price et al, 2018). The total area of Norway in this metric would amount to roughly, 1150 GW. Most area (580 GW or 50%) is available for onshore wind in the least constrained scenario (**None**). In the **Low** scenario, 265 GW or 23% of the land area and in the most constrained scenario (**Nature,Fauna,Sámi,Neigh**) less than 4 GW (corresponding to 0.0035% of the land area) can be installed.

# 3. Results

The key metrics we consider, to evaluate the impacts of the different onshore wind land use scenarios on electricity generation and storage equipment deployment in Norway, are: the total cost of the resulting cost-optimal electricity system, its design in terms of generation and storage capacities and their spatial deployment patterns.

## 3.1. Electricity system costs

The scenario **None**, which applies only technical restrictions, serves as a baseline to which the other scenarios are compared to. **None** applies no land use restrictions for the four studied dimensions. In it, the total system costs are close to €5 billion. The largest component of these costs is generation upkeep (i.e. fixed and variable operations and maintenance costs), primarily due to the large existing stock of hydroelectric power plants, whose capital costs are not included.

Figure TSCABS breaks the total system cost down into components for all spatial restriction scenarios considered. In the most restrictive scenario (**Nature,Fauna,Sámi,Neigh**) all remaining cost components are overshadowed by the load shedding costs. This happens because the level **High** of the *Neigh* dimension allows for very little wind power generation investment. Lacking alternative expandable generation sources, the model is forced to shed some load, which comes at a cost. This indicates the significant societal impacts of this scenario. The reduction in generation investment, which also leads to reduced generation upkeep costs and combined make up €1 billion, do not offset this.

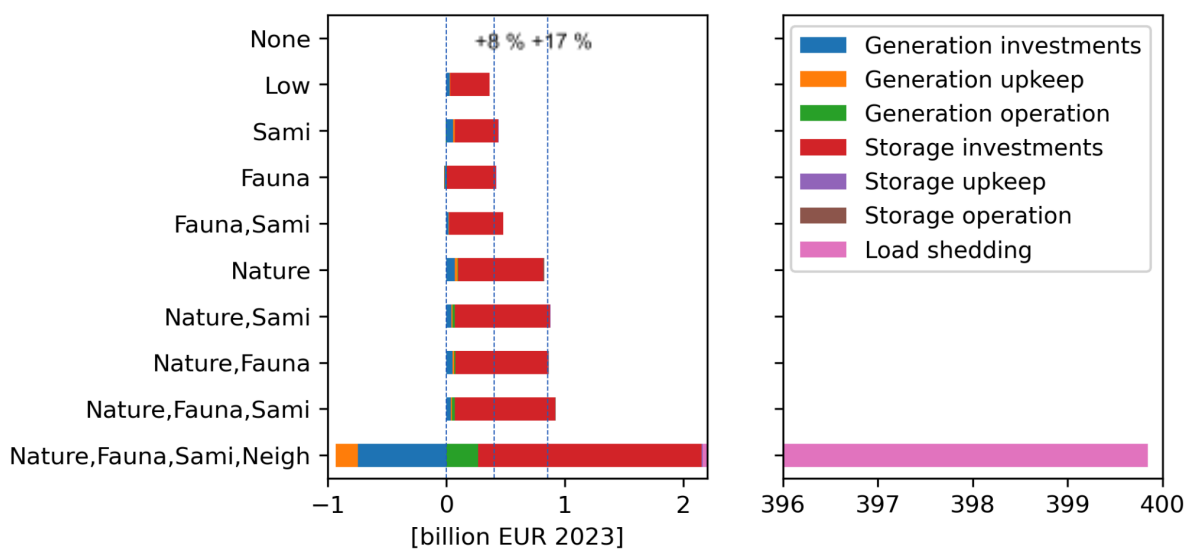

**Figure TSCABS:** Change of each component of total system costs per scenario compared to the cheapest scenario plotted. Markers at 0 %, 8 % and 17 % cost increase illustrate levels of similar cost increase between scenarios.

Figure TSCABS shows the cost increase of each component of the total system costs compared to the cheapest scenario **None,** in absolute terms. The largest increase stems

from storage investments in all but the **most restrictive scenario** where the cost of load shedding increases from negligible to almost €398 billion, making it the component increasing the most. Storage investment also more than doubles in scenarios with a High value for the nature dimension.

To understand the monetary costs to society for imposing the land use restrictions in the different scenarios, figure TSCABS also shows how much more expensive (percentage wise) the scenarios are compared to the baseline scenario, **None**. Three categories form: scenarios with a cost increase of around 8 % (~€400 million), scenarios with an increase of roughly 17 % (~€860 million), and a scenario with an increase of 8007 % (€398.9 billion) due to prohibitively expensive load shedding.

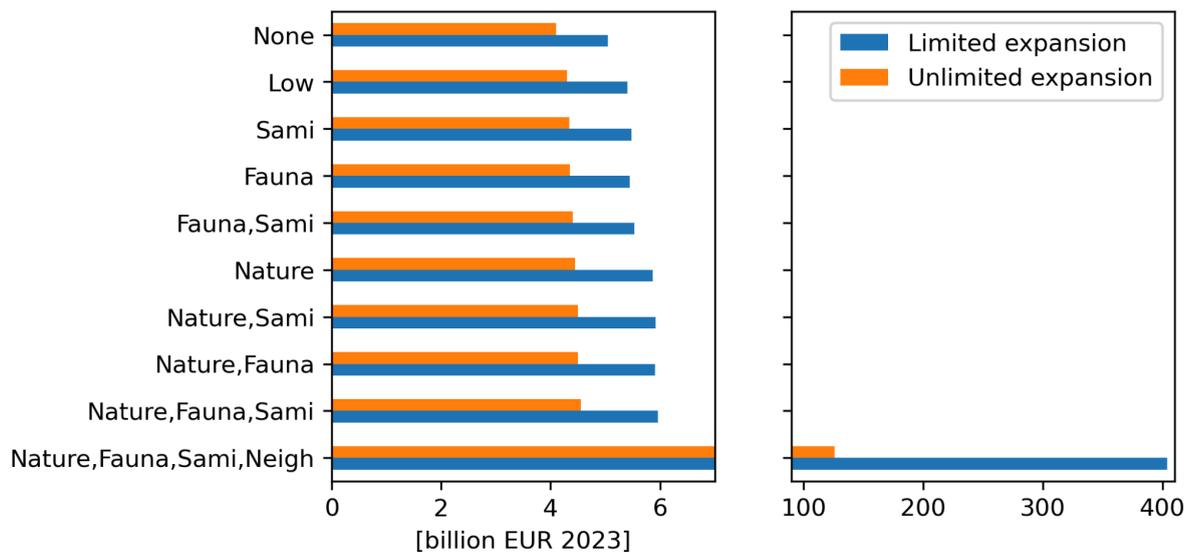

**Figure COSTSEXP:** Total system cost when allowing for grid expansion vs. when running with projected 2030 grid.

The figure COSTSEXP illustrates the system cost impacts of the assumption we make about restrictions on the expansion (depicted in blue) of the transmission grid in Norway. In orange, we can see what the system would cost if these assumptions were relaxed and investment into transmission expansion was possible. Total system cost levels would be lower by up to 30 % (€1.4 billion) in most cases under unrestricted transmission expansion. Interestingly, we see that the substantial system costs in the most restrictive case have markedly reduced, demonstrating the immense value offered by extra flexibility when the system design is highly constrained.

## 3.2. Electricity system capacities

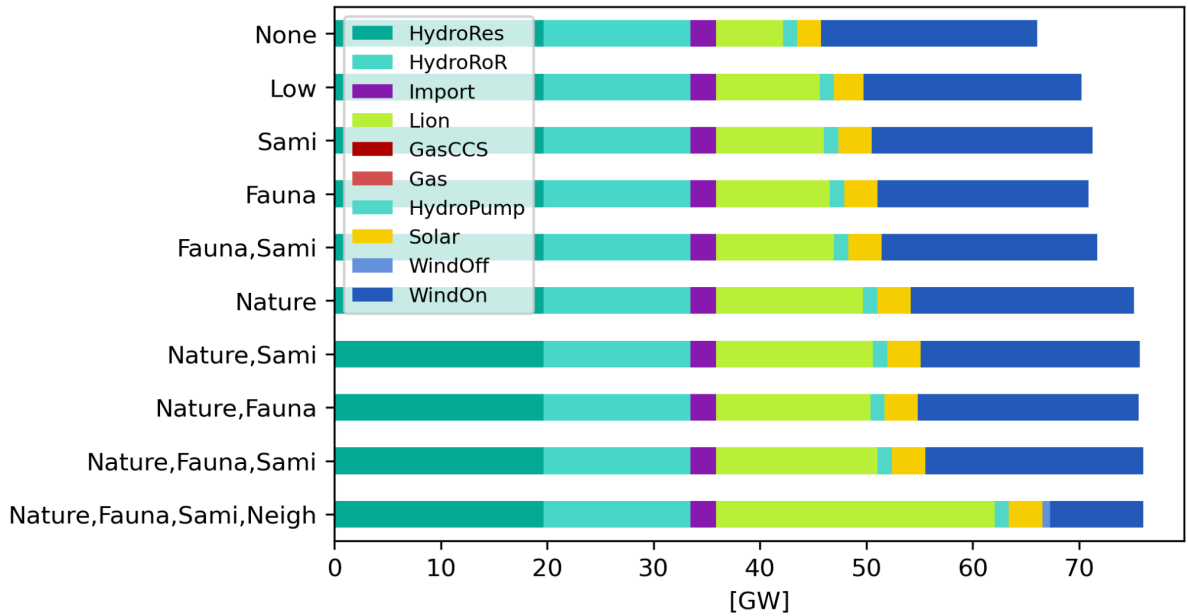

**Figure CAPGENSTOR:** Norwegian installed generation and storage power capacity

Figure CAPGENSTOR shows that with increasing limitations on the onshore wind deployment areas, battery storage capacity (Lion) increases. Note that hydropower capacity is pre-existing and, like import capacity, it can not be expanded.

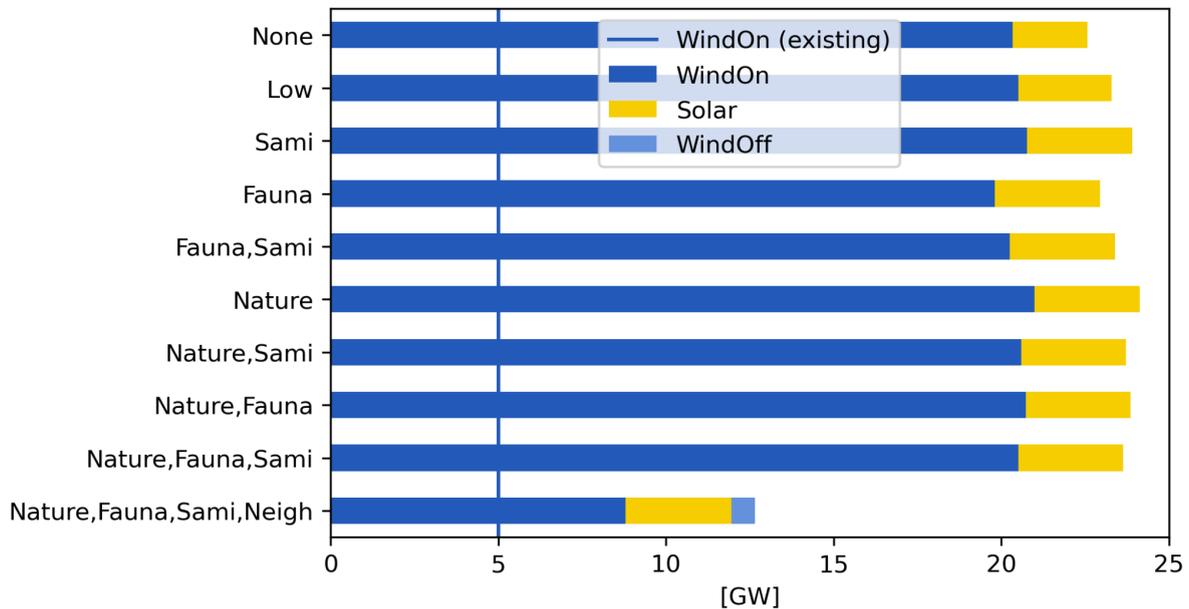

**Figure CAPTOT:** Norwegian optimal installed electricity generation capacity without hydropower and imports because they are fixed 2030. The vertical line depicts the existing onshore wind capacity.

Figure CAPTOT shows that offshore wind power generation does not play a role in the designed energy system, except for a small amount in the most restrictive scenario. There is little variance in the results, with the usual exception of the **most restrictive** scenario.

The reason there is variance at all is that the areas available to onshore wind change between scenarios, the installed capacity of it changes too. This means that the investments have to be slightly readjusted which either means the model has to substitute certain technologies in different regions (while keeping transmission in mind) or it has to turn to areas with slightly lower capacity factors, in turn increasing capacities. Still, as there is more area available than needed (see Figure AREABS), many promising sites are still available in all scenarios except **Nature,Fauna,Sami,Neigh**.

In the most constrained scenario, for onshore wind and solar PV, all available area is being used. Due to bottom-mounted offshore wind having around twice the capital costs of onshore wind, it is outcompeted in all scenarios but the most constrained one. Due to limited transmission line capacity between the county where it makes landfall (Rogaland) and its neighbouring counties, the model can not use the full offshore wind potential of 3 GW and only invests in 0.7 GW.

## 3.3. Spatial deployment of capacities

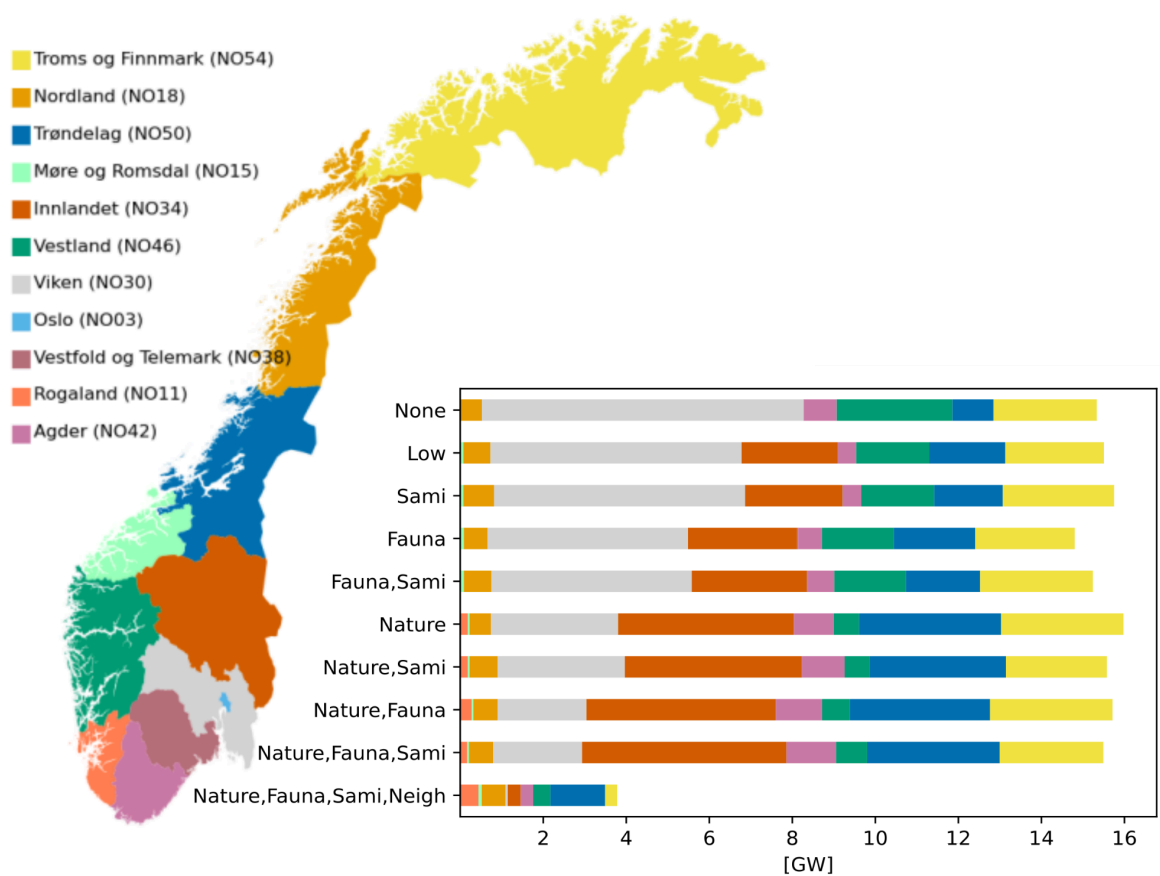

**Figure CAPNEWZ:** Spatial distribution of current (real) installed onshore wind generation capacity in GW [same as scenario Nature,Fauna,Sami,Neigh] and newly installed onshore wind generation capacity in GW per onshore wind restriction scenario

Figure CAPNEWZ shows the amount of wind power *newly* installed in the eleven Norwegian counties and due to the model implementation the amount of currently installed wind power

in the last scenario, as the model has to "rebuild" the installed capacity. Even though the total amount of newly installed wind is fairly constant in almost all the scenarios, additional land use restrictions, force the model to shift the new wind installations into different counties. That is because locations with the best wind conditions are not available any more under progressively more restrictive land use scenarios, for which the model compensates by building more capacity in less windy regions. The model then generates more from solar energy combined with battery storage. This leads to increasing deployment in the south-eastern county "Innlandet" (NO34). In general, the main regions that the model considers are "Viken" (NO30), "Troms og Finnmark" (NO54), "Vestland (NO46)", "Innlandet (NO34)" and "Trøndelag" (NO50). The only scenario where the model does not install wind energy in Innlandet is None. In the other scenarios the model does not have enough land area and access to good capacity factors in the other regions and as a result it deploys more and more wind energy in Innlandet. Through the level High in the Nature dimensions, Innlandet ascends to the top three regions for new onshore wind.

This figure also hints at why the model only chooses to invest in offshore wind in the most constrained scenario. There is almost no new onshore wind installed in Rogaland (NO11) (where most offshore wind could be built), as the region is self-sufficient on hydropower/imports in this model. This is due to the constrained transmission grid expansion, which leads to a bottleneck, so any additional generation in Rogaland cannot be moved out of the region.

Amongst others, limitations of this study are the simplified modelling of neighbouring countries and imports, the simplified modelling of demand side response (through load shedding), the fact that only Norwegian geodata is used for excluding areas. Especially in the High levels of many dimensions, buffers are applied, which may impact Norwegian territory, even if the source of the exclusion does not lie in Norway. Future research could use participatory approaches (e.g. workshops with stakeholders) to co-design the exclusion/restriction scenarios.

# 4. Discussion and conclusions

In line with its climate goals, Norway's electricity demand is rising as a result of the electrification of transportation and industry. There are also efforts to diversify the economy away from oil and gas by fostering the growth of electricity-intensive industries. Norway is currently relying mostly on hydropower (30GW) but the development potential is minimal. Over the last 10 years onshore wind has been expanded to 5 GW; however due to strong opposition there have been no new projects built over the last 3 years. We explore the implications of key social and environmental dimensions, shaping the future deployment of onshore wind, on the costs and design of electricity systems for Norway in 2030.

The most constraining scenario, which effectively means that wind energy is not visible from anyone's house (using a distance of 50 km between houses and wind turbines) reduces the possible maximum installed capacity for onshore wind from 580 GW to 4 GW. While this may look extreme, current opposition towards onshore wind can make such a scenario reality. The result of this heavily constrained scenario, which opts to prioritize minimizing the social and environmental impacts of onshore wind deployment, is that load needs to be shed. So the most restrictive scenario puts a high priority on local environmental protection, but that may impact Norway's efforts to mitigate climate change. This shed load is assumed to be highly costly, as it captures the socio-economic implications of not being able to serve demand. It means that the system is no longer reliable and in practice could mean leaving emissions targets behind, phasing out existing electricity-intensive industries, adapting behaviour, or importing energy. Imports are tight already, they reduce energy security and also increase emissions in practice as electricity from the rest of Europe is more $CO_2$ intensive and thus trade-offs between unpopular wind power and unpopular alternatives need to be considered. This situation is complicated by the layout and size of the grid in Norway and the fact that it is not possible to expand it beyond the plans in implementation until 2030. Relaxing any of those assumptions allows for cheaper system designs, emphasizing the significant advantages of incorporating extra flexibility when the system design is tightly constrained.

Strong protection of nature leads to an increase in total system costs of about 9 percentage points in costs, as well as an increase in installed battery storage capacity. There is also a noticeable shift of the location of newly installed wind capacity from populous Viken to Innlandet. From the spatial perspective, the cost-optimal solution concentrates wind power in the windiest counties and close to demand, which is in the South of Norway as we assume transmission to be fixed to current limits. With increasing land use restrictions, wind power installation is forced to spread out to less windy regions further away from demand, combined with storage. When the model can invest in transmission expansion, total system costs are lowered even in the most stringent scenario.

Concluding from this, Norway will need more flexibility due to rising electricity demand. This is in part due to the planned electrification of oil platforms and a general diversification away from oil to other industries that are often energy intensive. By 2030 this flexibility could come from different sources: increased transmission capacity (as seen in the transmission expansion cases), increased wind generation capacity (as seen in most cases of this study, increased storage capacity (as seen in all cases of this study), increased import volumes/capacity, demand side flexibility/reduction, reduced electrification.

However, each of these choices will lead to different costs and socio-environmental challenges. What is evident is that restricting wind energy will likely make the electricity system more costly. Opting not to build new wind and limiting transmission expansion can lead to blackouts resulting in large societal impacts (modelled here as up to €400 billion per year based on the lost load assumption in our modelling). In reality, this could mean missing climate targets or failing to meet the objectives of industrial policy (the additional demand would not be allowed to come online after all (e.g. industry, oil & gas electrification)). The central insight of this analysis is that the strict level **High** of the Neighbour dimension is completely incompatible with projected electrification. That is, some onshore wind will be needed one way or the other in order to meet demand by 2030, and the **High** level of the Neighbour dimension does not leave enough land for that. Yet, when reducing the restrictions on onshore wind or allowing for in-country transmission expansion, feasible system designs at a small fraction of that cost can be found. As such, people might have to accept wind power closer to their homes than they would like.

We are nearly five years away from 2030 and any infrastructure that we require to be online by then needs to be decided on today and implemented rapidly. Our analysis can help policymakers, regulators and the public make informed compromises as all options come at a monetary and socio-environmental cost. However, if no informed decision is made today, costs are likely to be very high.

## 5. Acknowledgements

We thank two summer students for their work on the exclusions and UiO:Energy for their funding.

# 6. Sources

Aasland, T. (2023, October 24). *The Minister of Petroleum and Energy's speech at the Energy Conference between Norway and the EU* [Taleartikkel]. Regjeringen.No; regjeringen.no. https://www.regjeringen.no/en/aktuelt/The-Minister-of-Petroleum-and-Energys-speech-at-the-Energy-conference-between-Norway-and-the-EU/id3004542/

Beyonder. (n.d.). *About Beyonder*. Beyonder. Retrieved 15 December 2023, from https://www.beyonder.no/facilities

Bolwig, S., Bolkesjø, T. F., Klitkou, A., Lund, P. D., Bergaentzlé, C., Borch, K., Olsen, O. J., Kirkerud, J. G., Chen, Y., Gunkel, P. A., & Skytte, K. (2020). Climate-friendly but socially rejected energy-transition pathways: The integration of techno-economic and socio-technical approaches in the Nordic-Baltic region. *Energy Research & Social Science*, *67*, 101559. https://doi.org/10.1016/j.erss.2020.101559

Buljan, A. (2022, August 19). TotalEnergies, Iberdrola and Norsk Havvind Now Jointly Known as 'Skjoldblad' in Norway. *Offshore Wind*. https://www.offshorewind.biz/2022/08/19/totalenergies-iberdrola-and-norsk-havvind-now-jointly-known-as-skjoldblad-in-norway/

Cheng, C., van Greevenbroek, K., & Viole, I. (2024). The competitive edge of Norway's hydrogen by 2030: Socio-environmental considerations. *International Journal of Hydrogen Energy*, *85*, 962–975. https://doi.org/10.1016/j.ijhydene.2024.08.377

Climate Analytics & NewClimate Institute. (2024, November 21). *The Climate Action Tracker—EU*. EU | Climate Action Tracker. https://climateactiontracker.org/countries/eu/

DeCarolis, J., Daly, H., Dodds, P., Keppo, I., Li, F., McDowall, W., Pye, S., Strachan, N., Trutnevyte, E., Usher, W., Winning, M., Yeh, S., & Zeyringer, M. (2017). Formalizing best practice for energy system optimization modelling. *Applied Energy*, *194*, 184–198. https://doi.org/10.1016/j.apenergy.2017.03.001

Egging, R., & Tomasgard, A. (2018). Norway's role in the European energy transition. *Energy Strategy Reviews*, *20*, 99–101. https://doi.org/10.1016/j.esr.2018.02.004

Elinor Batteries. (2023, January 11). *Skal bygge batterifabrikk i Trøndelag*. https://www.elinorbatteries.no/nyheter/elinor-batteries-lanserer-planer-om-batterifabrikk

Enevoldsen, P., & Permien, F.-H. (2018). Mapping the Wind Energy Potential of Sweden: A Sociotechnical Wind Atlas. *Journal of Renewable Energy*, *2018*(1), 1650794. https://doi.org/10.1155/2018/1650794

Fouche, G., & Klesty, V. (2023, March 1). Greta Thunberg detained by Norway police during pro-Sami protest. *Reuters*. https://www.reuters.com/world/europe/greta-thunberg-detained-by-norway-police-during-demonstration-2023-03-01/

FREYR Battery. (2022, June 29). *FREYR Battery Sanctions Construction of its Inaugural Gigafactory*. https://ir.freyrbattery.com/ir-news/press-releases/news-details/2022/FREYR-Battery-Sanctions-Construction-of-its-Inaugural-Gigafactory/default.aspx

Gambhir, A. (2019). Planning a Low-Carbon Energy Transition: What Can and Can't the Models Tell Us? *Joule*, *3*(8), 1795–1798. https://doi.org/10.1016/j.joule.2019.07.016

Gilad, D., Borgelt, J., May, R., Dorber, M., & Verones, F. (2024). Biodiversity impacts of Norway's renewable electricity grid. *Journal of Cleaner Production*, *469*, 143096. https://doi.org/10.1016/j.jclepro.2024.143096

Grimsrud, K., Hagem, C., Haaskjold, K., Lindhjem, H., & Nowell, M. (2023). Spatial Trade-Offs in National Land-Based Wind Power Production in Times of Biodiversity and Climate Crises. *Environmental and Resource Economics*. https://doi.org/10.1007/s10640-023-00764-8

Grochowicz, A., van Greevenbroek, K., Benth, F. E., & Zeyringer, M. (2023). Intersecting near-optimal spaces: European power systems with more resilience to weather variability. *Energy Economics*, *118*, 106496.
https://doi.org/10.1016/j.eneco.2022.106496

Gulbrandsen, L. H., Inderberg, T. H. J., & Jevnaker, T. (2021). Is political steering gone with the wind? Administrative power and wind energy licensing practices in Norway. *Energy Research & Social Science*, *74*, 101963.
https://doi.org/10.1016/j.erss.2021.101963

Hanna, R., & Gross, R. (2021). How do energy systems model and scenario studies explicitly represent socio-economic, political and technological disruption and discontinuity? Implications for policy and practitioners. *Energy Policy*, *149*, 111984.
https://doi.org/10.1016/j.enpol.2020.111984

Hansson, J., Unluturk, B., Wisell, T., Chang, M., Haaskjold, K., Hvidsten, T. V., Zeyringer, M., Hasager, M., Vested, M. H., Babri, S., Pinel, D., & Hjelkrem, O. A. (2023). *Fossil-free and resource efficient transport—Nordic Energy Outlooks—Final report WP4*. SINTEF Energi AS. https://sintef.brage.unit.no/sintef-xmlui/handle/11250/3115441

Hersbach, H., Bell, B., Berrisford, P., Biavati, G., Horányi, A., Muñoz Sabater, J., Nicolas, J., Peubey, C., Radu, R., Rozum, I., Schepers, D., Simmons, A., Soci, C., Dee, D., & Thépaut, J-N. (2018). *ERA5 hourly data on single levels from 1940 to present* [Dataset]. Copernicus Climate Change Service (C3S) Climate Data Store (CDS).
https://doi.org/10.24381/CDS.ADBB2D47

Hirt, L. F., Schell, G., Sahakian, M., & Trutnevyte, E. (2020). A review of linking models and socio-technical transitions theories for energy and climate solutions. *Environmental Innovation and Societal Transitions*, *35*, 162–179.
https://doi.org/10.1016/j.eist.2020.03.002

Hofmann, F., Hampp, J., Neumann, F., Brown, T., & Hörsch, J. (2021). atlite: A Lightweight Python Package for Calculating Renewable Power Potentials and Time Series. *Journal of Open Source Software*, *6*(62), 3294. https://doi.org/10.21105/joss.03294

Höltinger, S., Salak, B., Schauppenlehner, T., Scherhaufer, P., & Schmidt, J. (2016). Austria's

wind energy potential – A participatory modeling approach to assess socio-political and market acceptance. *Energy Policy*, *98*, 49–61. https://doi.org/10.1016/j.enpol.2016.08.010

Hovland, K. M. (2022, December 20). *Equinor trenger mye kraft i nord: – Støvsuger landsdelen*. https://e24.no/i/vekEvm

Inderberg, T. H. J., Nykamp, H. A., Olkkonen, V., Rosenberg, E., & Taranger, K. K. (2024). Identifying and analysing important model assumptions: Combining techno-economic and political feasibility of deep decarbonisation pathways in Norway. *Energy Research & Social Science*, *112*, 103496. https://doi.org/10.1016/j.erss.2024.103496

IPCC. (2023). *Climate Change 2022 – Impacts, Adaptation and Vulnerability: Working Group II Contribution to the Sixth Assessment Report of the Intergovernmental Panel on Climate Change* (1st ed.). Cambridge University Press. https://doi.org/10.1017/9781009325844

Karlstrøm, H., & Ryghaug, M. (2014). Public attitudes towards renewable energy technologies in Norway. The role of party preferences. *Energy Policy*, *67*, 656–663. https://doi.org/10.1016/j.enpol.2013.11.049

Katinas, V., Marčiukaitis, M., & Tamašauskienė, M. (2016). Analysis of the wind turbine noise emissions and impact on the environment. *Renewable and Sustainable Energy Reviews*, *58*, 825–831. https://doi.org/10.1016/j.rser.2015.12.140

Krumm, A., Süsser, D., & Blechinger, P. (2022). Modelling social aspects of the energy transition: What is the current representation of social factors in energy models? *Energy*, *239*, 121706. https://doi.org/10.1016/j.energy.2021.121706

Kurland, S. D. (2019). Energy use for GWh-scale lithium-ion battery production. *Environmental Research Communications*, *2*(1), 012001. https://doi.org/10.1088/2515-7620/ab5e1e

Lange, N. L., & Pochhammer, J. (2024, September 23). *Offshore wind in Germany and Europe- status quo and outlook*.

https://www.taylorwessing.com/en/insights-and-events/insights/2024/08/offshore-wind-in-deutschland-und-europa

Lawrence, R. (2014). Internal Colonisation and Indigenous Resource Sovereignty: Wind Power Developments on Traditional Saami Lands. *Environment and Planning D: Society and Space*, *32*(6), 1036–1053. https://doi.org/10.1068/d9012

McKenna, R., Hollnaicher, S., & Fichtner, W. (2014). Cost-potential curves for onshore wind energy: A high-resolution analysis for Germany. *Applied Energy*, *115*, 103–115. https://doi.org/10.1016/j.apenergy.2013.10.030

Meld. St. 13 (2020–2021), 232 (2021). https://www.regjeringen.no/contentassets/a78ecf5ad2344fa5ae4a394412ef8975/en-gb/pdfs/stm202020210013000engpdfs.pdf

Meld. St. 28 (2019-2020), Vindkraft På Land-Endringer i Konsesjonsbehandlingen 90 (2020). https://www.regjeringen.no/contentassets/f89e946defa24e57aaeb6bd25d949b7b/no/pdfs/stm201920200028000dddpdfs.pdf

Ministry of Climate and Environment. (2015, December 18). *Meld. St. 14 (2015–2016)* [Stortingsmelding]. Government.No; regjeringen.no. https://www.regjeringen.no/en/dokumenter/meld.-st.-14-20152016/id2468099/

Ministry of Climate and Environment. (2021). *Retningslinje for behandling av støy i arealplanlegging* [Retningslinjer]. regjeringen.no. https://www.regjeringen.no/no/dokumenter/retningslinje-for-behandling-av-stoy-i-arealplanlegging/id2857574/

Mölder, F., Jablonski, K. P., Letcher, B., Hall, M. B., Tomkins-Tinch, C. H., Sochat, V., Forster, J., Lee, S., Twardziok, S. O., Kanitz, A., Wilm, A., Holtgrewe, M., Rahmann, S., Nahnsen, S., & Köster, J. (2021). *Sustainable data analysis with Snakemake* (No. 10:33). F1000Research. https://doi.org/10.12688/f1000research.29032.2

Moore, A., Price, J., & Zeyringer, M. (2018). The role of floating offshore wind in a renewable focused electricity system for Great Britain in 2050. *Energy Strategy Reviews*, *22*,

270–278. Scopus. https://doi.org/10.1016/j.esr.2018.10.002

Morrow Batteries. (2023). *About Us*. https://www.morrowbatteries.com/about-us

Nikas, A., Lieu, J., Sorman, A., Gambhir, A., Turhan, E., Baptista, B. V., & Doukas, H. (2020). The desirability of transitions in demand: Incorporating behavioural and societal transformations into energy modelling. *Energy Research & Social Science*, *70*, 101780. https://doi.org/10.1016/j.erss.2020.101780

Norderhaug, M. (2023, March 9). *Energistatistikk for norske kommuner og fylker*. https://viken.no/tjenester/planlegging/analyse-statistikk-og-kart/analyse-statistikk-og-kart-for-viken/aktuelt-statistikk-og-kart/energistatistikk-for-norske-kommuner-og-fylker.75237.aspx

Norwegian Environment Agency. (2019a). *Faggrunnlag – Landskap—Underlagsdokument til nasjonal ramme for vindkraft* (No. M–1312). https://www.miljodirektoratet.no/globalassets/publikasjoner/m1312/m1312.pdf

Norwegian Environment Agency. (2019b). *Faggrunnlag – Naturtyper—Underlagsdokument til nasjonal ramme for vindkraft* (No. M–1311). https://www.miljodirektoratet.no/globalassets/publikasjoner/m1311/m1311.pdf

Norwegian Water Resources and Energy Directorate. (2012). *Landskap, friluftsliv og reiseliv – fagrapport til strategisk konsekvensutredning av fornybar energiproduksjon til havs*. https://publikasjoner.nve.no/rapport/2012/rapport2012_54.pdf

Norwegian Water Resources and Energy Directorate. (2013). *Offshore wind power in Norway—Strategic environmental assessment* (No. Rapport 47-12). https://publikasjoner.nve.no/diverse/2013/havvindsummary2013.pdf

Norwegian Water Resources and Energy Directorate. (2018). *Nasjonal ramme for vindkraft—Temarapport om nabovirkninger* (No. Nr 72/2018). Norwegian Water Resources and Energy Directorate. https://publikasjoner.nve.no/rapport/2018/rapport2018_72.pdf

Norwegian Water Resources and Energy Directorate. (2019a). *Forslag til nasjonal ramme for vindkraft. http://publikasjoner.nve.no/rapport/2019/rapport2019_12.pdf

Norwegian Water Resources and Energy Directorate. (2019b). *Vindhastighet*. NVE Nasjonal ramme for vindkraft. https://temakart.nve.no/

Norwegian Water Resources and Energy Directorate. (2023, June 26). *Produksjonsrapporter*. Produksjonsrapporter - NVE. https://www.nve.no/energi/energisystem/vindkraft-paa-land/produksjonsrapporter/

Outdoor Recreation Act, LOV 1957-06-28 (1957). https://www.regjeringen.no/en/dokumenter/outdoor-recreation-act/id172932/

Ovaere, M., Heylen, E., Proost, S., Deconinck, G., & Van Hertem, D. (2016). *How Detailed Value of Lost Load Data Impact Power System Reliability Decisions: A Trade-Off between Efficiency and Equity* (SSRN Scholarly Paper No. 2877129). Social Science Research Network. https://doi.org/10.2139/ssrn.2877129

Permien, F.-H., & Enevoldsen, P. (2019). Socio-technical constraints in German wind power planning: An example of the failed interdisciplinary challenge for academia. *Energy Research & Social Science*, *55*, 122–133. https://doi.org/10.1016/j.erss.2019.04.021

Pfenninger, S., Hawkes, A., & Keirstead, J. (2014). Energy systems modeling for twenty-first century energy challenges. *Renewable and Sustainable Energy Reviews*, *33*, 74–86. https://doi.org/10.1016/j.rser.2014.02.003

Price, J., Keppo, I., & Dodds, P. E. (2023). The role of new nuclear power in the UK's net-zero emissions energy system. *Energy*, *262*. Scopus. https://doi.org/10.1016/j.energy.2022.125450

Price, J., Mainzer, K., Petrovic, S., Zeyringer, M., & McKenna, R. (2022). The Implications of Landscape Visual Impact on Future Highly Renewable Power Systems: A Case Study for Great Britain. *IEEE Transactions on Power Systems*, *37*(4), 3311–3320. Scopus. https://doi.org/10.1109/TPWRS.2020.2992061

Price, J., & Zeyringer, M. (2022). highRES-Europe: The high spatial and temporal Resolution Electricity System model for Europe. *SoftwareX*, *17*, 101003. Scopus.
27

28
https://doi.org/10.1016/j.softx.2022.101003

Price, J., Zeyringer, M., Konadu, D., Mourao, Z. S., Moore, A., & Sharp, E. (2018). Low carbon electricity systems for Great Britain in 2050: An energy-land-water perspective. *Applied Energy*, *228*, 928–941. Scopus. https://doi.org/10.1016/j.apenergy.2018.06.127

Reindeer Herding Act, LOV-2007-06-15-40 (2021). https://lovdata.no/dokument/NL/lov/2007-06-15-40?q=LOV-2007-06-15-40

Rinne, E., Holttinen, H., Kiviluoma, J., & Rissanen, S. (2018). Effects of turbine technology and land use on wind power resource potential. *Nature Energy*, *3*(6), 494–500. https://doi.org/10.1038/s41560-018-0137-9

Rydell, J., Engström, H., Hedenström, A., Larsen, J. K., Pettersson, J., & Green, M. (2012). *The effect of wind power on birds and bats – A synthesis* (No. 6511).

Sørensen, Å. L., Westad, M. C., Delgado, B. M., & Lindberg, K. B. (2022). Stochastic load profile generator for residential EV charging. *E3S Web of Conferences*, *362*, 03005. https://doi.org/10.1051/e3sconf/202236203005

Spilde, D., Hole, J., Haukeli, I. E., Haug, M., & Brunvoll, A. T. (2020). *Elektrifisering av landbaserte industrianlegg i Norge*. NVE. https://publikasjoner.nve.no/rapport/2020/rapport2020_18.pdf

Statistics Norway. (2024, November 19). *Electricity*. Electricity | SSB. https://www.ssb.no/en/energi-og-industri/energi/statistikk/elektrisitet

Statnett. (2023a). *Forbruksutvikling i Norge 2022-2050*. https://www.statnett.no/globalassets/for-aktorer-i-kraftsystemet/planer-og-analyser/lma/forbruksutvikling-i-norge-2022-2050---delrapport-til-lma-2022-2050.pdf

Statnett. (2023b). *Langsiktig markedsanalyse—Norge, Norden og Europa 2022-2050*. https://www.statnett.no/globalassets/for-aktorer-i-kraftsystemet/planer-og-analyser/lma/langsiktig-markedsanalyse-2022-2050.pdf

Statnett. (2024). *Kortsiktig Markedsanalyse 2024-2029*.

https://www.statnett.no/globalassets/for-aktorer-i-kraftsystemet/planer-og-analyser/kma/kortsiktig-markedsanalyse-2024-2029.pdf

Summary of Proposition No. 52 (2008-2009) to the Storting Concerning an Act Relating to the Management of Biological, Geological and Landscape Diversity (Nature Diversity Act) (2009).

https://www.regjeringen.no/en/dokumenter/ot.prp.-nr.-52-2008-2009/id552112/

Süsser, D., Martin, N., Stavrakas, V., Gaschnig, H., Talens-Peiró, L., Flamos, A., Madrid-López, C., & Lilliestam, J. (2022). Why energy models should integrate social and environmental factors: Assessing user needs, omission impacts, and real-word accuracy in the European Union. *Energy Research & Social Science*, *92*, 102775. https://doi.org/10.1016/j.erss.2022.102775

Trutnevyte, E. (2016). Does cost optimization approximate the real-world energy transition? *Energy*, *106*, 182–193. https://doi.org/10.1016/j.energy.2016.03.038

United Nations Environment Programme, Olhoff, A., Bataille, C., Christensen, J., Den Elzen, M., Fransen, T., Grant, N., Blok, K., Kejun, J., Soubeyran, E., Lamb, W., Levin, K., Portugal-Pereira, J., Pathak, M., Kuramochi, T., Strinati, C., Roe, S., & Rogelj, J. (2024). *Emissions Gap Report 2024: No more hot air … please! With a massive gap between rhetoric and reality, countries draft new climate commitments*. United Nations Environment Programme. https://doi.org/10.59117/20.500.11822/46404

Vogt, L. F., Tollersrud, T., Sørenes, A., & Solli, I. J. (2023, March 13). Sterke reaksjoner på kraftforbruk i nytt dataanlegg. *NRK*. https://www.nrk.no/innlandet/datasenter-vil-bruke-enorme-mengder-strom-_-lo-forbund-reagerer-1.16329778

Zeyringer, M., Fais, B., Keppo, I., & Price, J. (2018). The potential of marine energy technologies in the UK – Evaluation from a systems perspective. *Renewable Energy*, *115*, 1281–1293. Scopus. https://doi.org/10.1016/j.renene.2017.07.092

Zeyringer, M., Price, J., Fais, B., Li, P.-H., & Sharp, E. (2018). Designing low-carbon power

systems for Great Britain in 2050 that are robust to the spatiotemporal and inter-annual variability of weather. *Nature Energy*, *3*(5), 395–403. Scopus. https://doi.org/10.1038/s41560-018-0128-x

# 7. Appendix

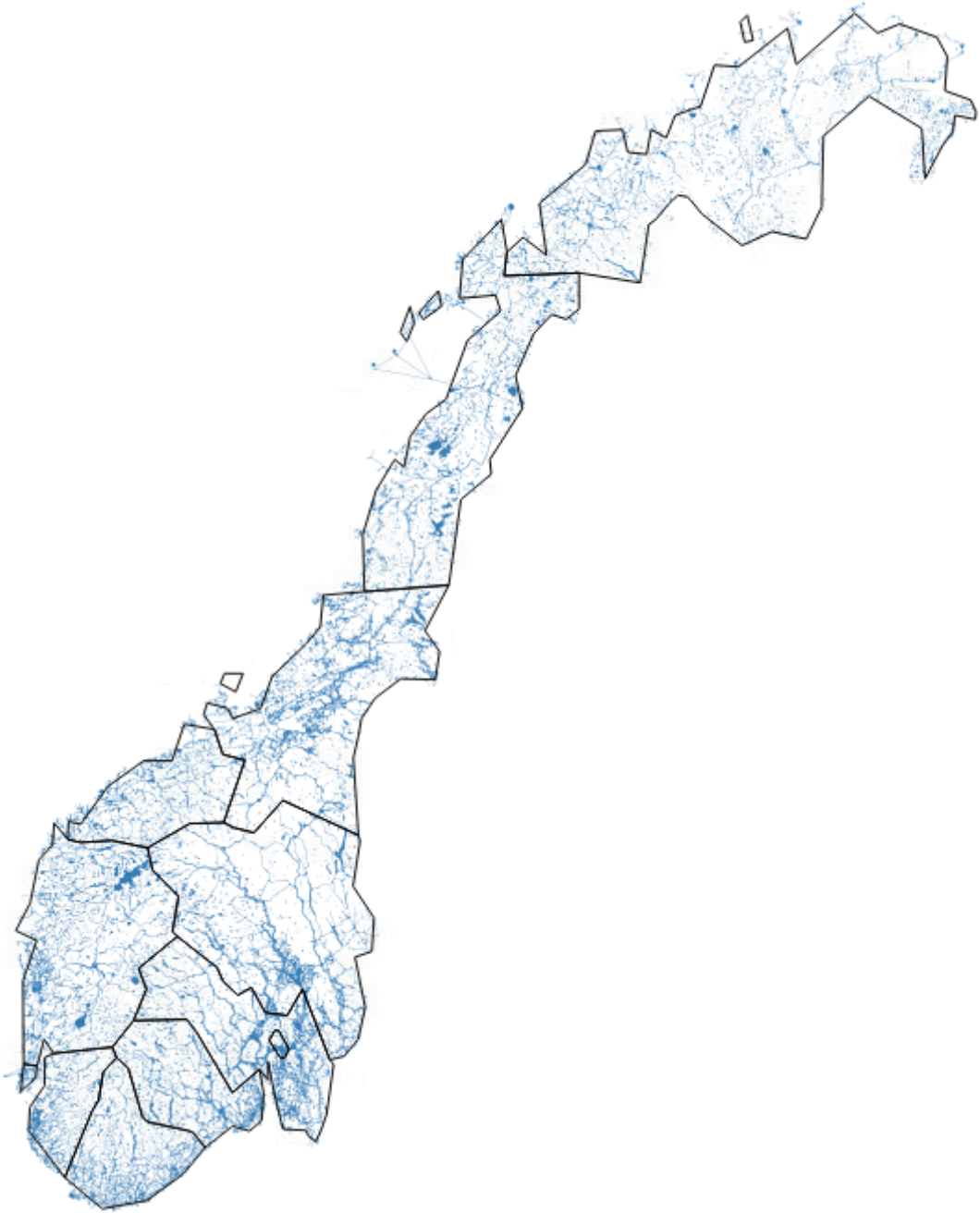

Figure TECEXCL: Technical restrictions

## 7.1. Case selection: Why Norway?

The Norwegian power system is facing an increase in electricity demand from the electrification of transport, heating and industry, while the traditional generation source (hydropower) is not able to meet all of this increase due to environmental limits.

This leads to concerns about a power deficit by 2030 (Statnett, 2024) which we choose as the target year of the analysis.

The most cost-effective new sources of electricity generation are wind and solar power. Norwegian geography allows for some of the best on- and offshore wind resources in Europe (Egging & Tomasgard, 2018; Karlstrøm & Ryghaug, 2014). Norway has developed a great deal of expertise in offshore technology through the oil and gas sector, and is in a good position to play a role in the offshore wind sector. Both Equinor (the Norwegian state-owned petroleum company) and the Norwegian energy production industry are involved in the development of the largest offshore wind farms in Europe and the United States, including bottom-mounted and floating turbines. Nonetheless, floating wind power is a technology that is still transitioning from the demonstration stage to full-scale testing, and therefore cannot be considered a key technology to meet 2030 decarbonization goals (emissions reduction by at least 50 % and towards 55% compared to 1990 levels) (Meld. St. 13 (2020–2021), 2021).

Furthermore, even though licence applications have been opened for offshore wind renewable energy production in 2021, on average, applications take an average of five-and-a-half years (Gulbrandsen et al., 2021). As a result, in Norway, onshore wind and solar power currently offer the greatest potential for new renewable energy production to reach its 2030 decarbonization goals, as is also the case for other countries in the Nordic region.

However, onshore wind energy projects in Norway are facing opposition from nature conservation and recreational groups, and local communities (Gulbrandsen et al., 2021; Karlstrøm & Ryghaug, 2014). Wind farms, as well as solar power often require large areas that can have an impact on connectivity areas for fauna and ecosystems, which may cause disturbance and habitat fragmentation for mammals, birds and other animals (Gilad et al., 2024).

In 2019, the NVE proposed a map of 13 major geographical areas regarded as most suitable for locating onshore wind power in Norway amounting to 29,000 km$^2$ (Norwegian Water Resources and Energy Directorate, 2019b). These roughly 9 % of the total land area of Norway could correspond to up to 290 gigawatts (GW) of onshore wind power, if fully used according to the density mentioned in the framework. In this map, the NVE placed emphasis on avoiding conflicts with protected natural areas, outdoor life, noise, birds and wildlife, cultural heritage and reindeer husbandry. However, after a round of consultations with municipalities, ministries, the Sámi[1] Parliament, nature conservation associations, and outdoor life associations among other organizations and individuals which were critical to the plan, the government decided to scrap the proposed framework (Gulbrandsen et al., 2021). Instead, the NVE has published a white paper on measures for tightening the processing of wind power concessions, where more consideration is given to impacts on landscapes, the environment, society, and neighbours (Meld. St. 28 (2019-2020), 2020).

---

[1] The Sámi are an indigenous people with traditional territories within the national borders of Finland, Norway, Sweden, and Russia

While renewable energy development, nature protection and social support may all be regarded as critical, an energy transition that balances all three is proving challenging in meeting Norway's future energy demand and achieving its decarbonization goals. For instance, the construction of wind farms has been considered by the Sámi Council as threatening the sustainability of reindeer herding (Lawrence, 2014). In 2021, Norway's supreme court ruled that two wind farms built at Fosen in central Norway violated Sámi human rights under international conventions. The future of these wind farms is still unclear. This uncertainty sparked protests in February 2023, where Sámi activists blocked the entrance to Norway's energy ministry, demanding the cease of operations of the energy plants (Fouche & Klesty, 2023). Therefore, socio-environmental constraints can have a large impact on the overall electricity generation capacity potential which will in turn affect optimal technology choices, system costs and the social feasibility of reaching the Paris Agreement.

While it is a socio-political decision to choose more expensive technologies, sites or mitigation options to minimize the socio-environmental impacts of VRE development, a spatially-explicit capacity assessment under different socio-environmental scenarios is missing to allow for such discussion. Here, we close this gap by performing a nationally specific analysis: we first study the NVE framework, previous licences, literature, and newspaper articles to design three scenarios of socio-environmental acceptance for onshore/offshore wind and solar energy. Based on the developed scenarios, we then conduct a GIS analysis to determine the spatially dependent capacity potential per technology and scenario.

The NVE's onshore wind power framework consists of an updated basis of knowledge and a map where 13 areas are regarded as most suitable for locating wind energy. These consider factors such as untouched nature, nature types, fauna, landscapes, outdoor life, cultural heritage, Sámi interests and populated areas (Norwegian Water Resources and Energy Directorate, 2019a). The framework is neither a development plan nor legally binding, but was meant to be a guiding document to select adequate areas to install wind farms. Net capacity and transmission grids are included in the framework.

The Energy Act regulates the planning of onshore energy production in Norway. The NVE does an assessment for every licence application based on whether the advantages of the project are higher than the effects on public and private interests. These interests include nature conservation, cultural heritage, landscape outdoor life, migration of birds, fish, Sámi interests and local communities.[2] As an example, nature's value is protected through the Nature Diversity Act[3] and shall ensure biological and geological diversity today and in the future.[4]

---

[2] Ot.prp.nr. 43 (1989-90) s. 84

[3] Act relating to the management of biological, geological and landscape diversity [Nature Diversity Act]

[4] Ot.prp.nr. 52 (2008-2009) page 371.

## 7.2. Model changes

Compared to the version employed by Price et al. (2018), we change the modelled battery technology from Sodium Sulphur to Lithium-ion batteries. We change the weather data from using the year 2006 to 2010, which we choose based on (Price et al., 2023). The land use constraints also differ and are described in Section 2.3. Water availability is not taken into account in this paper, as thermal generation is not very relevant in the Norwegian context.

The workflow management system snakemake (Mölder et al., 2021) is employed to automate the process of preparing the required input data, running the model and converting the model output for each of the scenarios. We conduct the analysis in Python using Jupyter notebooks.

Version 0.2.4 of atlite is used. A slight modification is added, to be able to extract the weather data on a grid cell level (of the reanalysis data), compared to the default which is to aggregate over time or space.

The possibility exists that some existing wind power was built in areas that would be excluded from wind power development in this analysis. We therefore first add the locations of existing wind power to the exclusion zones. To guarantee that wind can be built by the model in those zones where wind power exists, we calculate a second exclusion zone that excludes all area except for area that hosts currently installed plants. Next, we add both matrices containing the exclusions to result in a matrix that excludes the desired areas according to the scenario guidelines, but also includes (in the buildable zones) areas where wind already exists.

Since the exclusion amounts are measured in area (km²) but the existing wind power is quantified in installed power (MW) and the locations of existing parks are points without an area, we create a buffer around those points according to the assumed installation density (MW/km²) and the installed capacity at this point. We feed the resulting area from this process into the workflow described in the paragraph above.

## 7.3. Data description

Table CORINE:

| Incremental ID | Semantic ID | Top level category | Mid-level Category | Sublevel category |
|---|---|---|---|---|
| 2 | 112 | Artificial surfaces | Urban fabric | Discontinuous urban fabric |
| 4 | 122 | Artificial surfaces | Industrial, commercial and transport units | Road and rail networks and associated land |
| 5 | 123 | Artificial surfaces | Industrial, commercial and transport units | Port areas |
| 6 | 124 | Artificial surfaces | Industrial, commercial and transport units | Airports |
| 10 | 141 | Artificial surfaces | Artificial, non-agricultural vegetated areas | Green urban areas |
| 11 | 142 | Artificial surfaces | Artificial, | Sport and leisure facilities |

| | | | non-agricultural vegetated areas | |
|---|---|---|---|---|
| 12 | 211 | Agricultural areas | Arable land | Non-irrigated arable land |
| 13 | 212 | Agricultural areas | Arable land | Permanently irrigated land |
| 14 | 213 | Agricultural areas | Arable land | Rice fields |
| 15 | 221 | Agricultural areas | Permanent crops | Vineyards |
| 16 | 222 | Agricultural areas | Permanent crops | Fruit trees and berry plantations |
| 17 | 223 | Agricultural areas | Permanent crops | Olive groves |
| 18 | 231 | Agricultural areas | Pastures | Pastures |
| 19 | 241 | Agricultural areas | Heterogeneous agricultural areas | Annual crops associated with permanent crops |
| 20 | 242 | Agricultural areas | Heterogeneous agricultural areas | Complex cultivation patterns |
| 21 | 243 | Agricultural areas | Heterogeneous agricultural areas | Land principally occupied by agriculture, with significant areas of natural vegetation |
| 22 | 244 | Agricultural areas | Heterogeneous agricultural areas | Agro-forestry areas |
| 23 | 311 | Forest and semi natural areas | Forests | Broad-leaved forest |
| 24 | 312 | Forest and semi natural areas | Forests | Coniferous forest |
| 25 | 313 | Forest and semi natural areas | Forests | Mixed forest |
| 34 | 335 | Forest and semi natural areas | Open spaces with little or no vegetation | Glaciers and perpetual snow |
| 35 | 411 | Wetlands | Inland wetlands | Inland marshes |
| 36 | 412 | Wetlands | Inland wetlands | Peat bogs |
| 37 | 421 | Wetlands | Maritime wetlands | Salt marshes |
| 38 | 422 | Wetlands | Maritime wetlands | Salines |
| 39 | 423 | Wetlands | Maritime wetlands | Intertidal flats |
| 40 | 511 | Water bodies | Inland waters | Water courses |
| 41 | 512 | Water bodies | Inland waters | Water bodies |
| 42 | 521 | Water bodies | Marine waters | Coastal lagoons |
| 43 | 522 | Water bodies | Marine waters | Estuaries |
| 44 | 523 | Water bodies | Marine waters | Sea and ocean |

Table CAPLIM: Upper capacity limits per Norwegian county

| Zone | Import [GW] | PV [GW] |
|------|-------------|---------|
| NO03 | 0 | 0.281 |
| NO11 | 0.35 | 0.265 |
| NO15 | 0 | 0.076 |
| NO18 | 0.23125 | 0.024 |
| NO30 | 0.53 | 0.728 |
| NO34 | 0 | 0.276 |
| NO38 | 0 | 0.386 |
| NO42 | 0.7625 | 0.26 |
| NO46 | 0.35 | 0.218 |
| NO50 | 0.2 | 0.638 |
| NO54 | 0.03 | 0.007 |

Table TRANSCAP: Transmission capacities between Norwegian counties.

| Zone 1 | Zone 2 | Link capacity [MW] |
|--------|--------|--------------------|
| NO03 | NO30 | 3000 |
| NO11 | NO38 | 900 |
| NO11 | NO42 | 1200 |
| NO11 | NO46 | 750 |
| NO15 | NO34 | 500 |
| NO15 | NO46 | 3000 |
| NO15 | NO50 | 1350 |
| NO18 | NO50 | 1350 |
| NO18 | NO54 | 600 |
| NO30 | NO34 | 7000 |
| NO30 | NO38 | 500 |
| NO30 | NO46 | 3900 |
| NO34 | NO50 | 600 |
| NO38 | NO42 | 1200 |

## 7.4. Electricity demand

Electricity demand is modelled according to a consumption prognosis by the Norwegian transmission system operator, Statnett (2023a). The methodology is previously described in (Hansson et al., 2023) and only additional details follow here. The changes in electricity demand between 2022 and 2030 resulting from the prognosis is summarized in Table 8.

| Sector | 2022 [TWh] | 2030 [TWh] | Change [TWh] |
|---|---|---|---|
| Battery production and data centres | 1 | 6 | 5 |
| Petroleum | 9 | 20 | 11 |
| Industry | 47 | 59 | 12 |
| Electric transport | 3 | 13 | 10 |
| Other consumption | 80 | 79 | -1 |
| Total | 140 | 178 | 38 |

**Table 8:** Electricity consumption in 2022 and prognosis for 2030, from Statnett (2023a).

Historical demand at hourly resolution for 1951 to 2021 is scaled so that the average yearly demand equals 140 TWh and yearly variation is kept. The 10 TWh from electric transport is distributed temporally, using an electric vehicle load curve from Sørensen et al. (2022), and temporally as described in (Hansson et al., 2023). The 12 TWh from industry is distributed spatially based on each county's share of total energy use in industry from (Norderhaug, 2023). Of the 11 TWh from the petroleum sector, 4 TWh is allocated to electrification of onshore gas processing plants: 3.6 TWh in Troms og Finnmark (Hovland, 2022) and 0.4 TWh in Vestland (Spilde et al., 2020). The remaining 7 TWh is distributed evenly between twelve different offshore petroleum plants with potential for electrification: seven in Vestland, one in Møre og Romsdal, two in Trøndelag, and two in Troms og Finnmark (Statnett, 2023a).

The electricity consumption of battery production and data centres is according to Statnett's prognosis expected to increase by 5 TWh from 2022 to 2030 (Statnett, 2023a). However, we have identified planned battery factories that alone could make up more than 8 TWh of electricity demand in 2030. Four factories for battery production are planned in Norway, with annual production capacity of 43 GWh in Agder by 2028 (Morrow Batteries, 2023), 200 GWh (we assume 50 GWh as this capacity might not be realized by 2030) in Nordland by 2030 (FREYR Battery, 2022), 40 GWh in Trøndelag by 2030 (Elinor Batteries, 2023), and a factory in Rogaland with unknown production capacity (Beyonder, n.d.) (we assume 30 GWh by 2030). To go from the production capacity to electricity consumption, we apply an electricity use per GWh battery production of 50 GWh (Kurland, 2019). Demand from battery production is distributed spatially based on the identified factories. Of the 2 TWh from data centres, 1.3 TWh is allocated to Innlandet based on a planned data centre (Vogt et al., 2023) while the rest is evenly distributed between the other counties.

## 7.5. Geodata

| **TECHNICAL CONSTRAINTS** | |
|---|---|
| **VARIABLE** | **DATA SET** |
| Roads – 200m buffer | https://kartkatalog.geonorge.no/metadata/kartverket/vbase/96104f20-15f6-460e-a907-501a65e2f9ce<br><br>Title: Vbase<br><br>Format: Shape |
| Railway 200m buffer | https://kartkatalog.geonorge.no/metadata/bane-nor-sf/jernbane-banenettverk/c3da3591-cded-4584-a4b1-bc61b7d1f4f2<br><br>Title: Jernbane - Banenettverk<br><br>Format: GML |
| Airports 2 km buffer | https://land.copernicus.eu/pan-european/corine-land-cover/clc2018?tab=download<br><br>Select by attribute: clc18_kode = 124<br><br>Data explanation: https://land.copernicus.eu/user-corner/technical-library/corine-land-cover-nomenclature-guidelines/html/index-clc-124.html |
| Water courses | https://land.copernicus.eu/pan-european/corine-land-cover/clc2018?tab=download<br><br>Select by attribute: clc18_kode = 511<br><br>Data explanation: https://land.copernicus.eu/user-corner/technical-library/corine-land-cover-nomenclature-guidelines/html/index-clc-511.html |

| Water bodies | https://land.copernicus.eu/pan-european/corine-land-cover/clc2018?tab=download |
| --- | --- |
| | Select by attribute: clc18_kode = 512 |
| | Data explanation: https://land.copernicus.eu/user-corner/technical-library/corine-land-cover-nomenclature-guidelines/html/index-clc-512.html |
| Glaciers | https://land.copernicus.eu/pan-european/corine-land-cover/clc2018?tab=download |
| | Select by attribute: clc18_kode = 335 |
| | https://land.copernicus.eu/user-corner/technical-library/corine-land-cover-nomenclature-guidelines/html/index-clc-335.html |
| Gradients steeper than 20 deg | https://hoydedata.no/LaserInnsyn/ |
| | Click: Nedlasting -> Landsdekkende -> Velg UTM-sone 33 -> DTM50 |
| | Needed to merge the data |

| **NEIGHBOURS** | |
| --- | --- |
| **LOW EXCLUSION** | |
| **VARIABLE** | **DATA SET** |
| Buildings – 400m buffer | https://kartkatalog.geonorge.no/metadata/kartverket/n250-kartdata/442cae64-b447-478d-b384-545bc1d9ab48<br><br>Title: N250 Kartdata<br><br>Format: SOSI<br><br>Folder: Arealdekke<br><br>Select by attribute: OBJTYPE = BymessigBebyggelse |
| Agglomerations – 1 km buffer | https://kartkatalog.geonorge.no/metadata/kartverket/n250-kartdata/442cae64-b447-478d-b384-545bc1d9ab48<br><br>Title: N250 Kartdata<br><br>Format: SOSI<br><br>File: Arealdekke<br><br>Select by attribute: OBJTYPE = Tettbebyggelse |
| Walking routes, ski runs, biking routes other-routes - 20m buffer | https://kartkatalog.geonorge.no/metadata/uuid/d1422d17-6d95-4ef1-96ab-8af31744dd63<br><br>Title: Tur- og friluftsruter |
| Cultural heritage:<br>● Sikringssoner<br>● Brannsmitteområder<br>● Lokaliteter | https://kartkatalog.geonorge.no/metadata/riksantikvaren/kulturminner-sikringssoner/0a3251bb-2a50-45d3-8674-58bade2fe673<br><br>Title: Kulturminner – Sikringssoner<br><br>Fortmat: FGDB<br><br>https://kartkatalog.geonorge.no/metadata/riksantikvaren/kulturminner-brannsmitteomrader/73f863ba-628f-48af-b7fa-30d3ab331b8d<br><br>Title: Kulturminner – Brannsmitteområder |

| | Format: FGDB https://kartkatalog.geonorge.no/metadata/riksantikvaren/kulturminner-lokaliteter/c6896f24-71f9-4203-9b6f-faf3bfe1f5ed  Title: Kulturminner – Lokaliteter  Format: FGDB |
|---|---|

| **FAUNA** |  |
|---|---|
| **LOW RESTRICTION** | |

| VARIABLE | DATASET |
|---|---|
| IBA – Important Bird Areas | Needed to request data from http://datazone.birdlife.org/site/requestgis |
| Very important nature for biodiversity | https://kartkatalog.miljodirektoratet.no/Dataset/Details/10  Chose Norge uten Svalbard and select viktige naturtyper.  Select by attribute: BMVERDI = S |

| **NATURE** |
|---|
| **LOW EXCLUSION** |

| VARIABLE | DATASET |
|---|---|
| Strict nature reserve – IUCN = 1a | https://kartkatalog.geonorge.no/metadata/uuid/5857ec0a-8d2c-4cd8-baa2-0dc54ae213b4<br><br>Title: Naturvernområder<br><br>Format: SOSI<br><br>Select by attribute: IUCN = 1 |

| **Sámi REINDEER HERDING** |  |
|---|---|
| **LOW EXCLUSION** | |
| **VARIABLE** | **DATASET** |
| Winter grazing land (Minimum grazing land) | https://kartkatalog.geonorge.no/metadata/landbruksdirektoratet/reindrift-arstidsbeite-vinterbeite/63f655ef-f625-43cf-a512-bb8164bf53a4<br><br>Title: Reindrift – Årstidsbeite – Vinterbeite<br><br>Format: SOSI |
| Calving area | https://kartkatalog.geonorge.no/metadata/landbruksdirektoratet/reindrift-arstidsbeite-varbeite/fa02a652-cd6d-4828-9fb5-7bd4515aa6d0<br><br>Title: Reindrift – Årstidsbeite – Vårbeite<br><br>Format: SOSI<br><br>Select by attribute: kodenavn = Vårbeite I |

| **PV** | |
|---|---|
| **LOW EXCLUSION** | |
| VARIABLE | DATASET |
| Land used for agriculture | https://kartkatalog.geonorge.no/metadata/norsk-institutt-for-biookonomi/arealressurskart-ar50-jordbruk/76255ebe-2a0e-401e-87c8-7618dd196cf2<br><br>Title: Arealressurskart – AR50 – Jordbruk<br><br>Format: SOSI, GML<br><br>SOSI: Akershus, Aust Agder, Buskerd, Finnmark, Hordaland, MøreRomsdalen, Oslo, Rogaland, Romsa Troms, Sogn Fjordane, Telemark, Vest Agder, Vestfold, Østfold<br><br>GML: Hedmark, Nordland, Oppland<br><br>Select by attribute: Artype = 20 |
| Very good soil quality | https://kartkatalog.geonorge.no/metadata/norsk-institutt-for-biookonomi/jordkvalitet/35c38144-c0a0-4ed9-a66f-21b80bc17fa7<br><br>Title: Jordkvalitet<br><br>Format: SOSI<br><br>Needed to download every region by itself and then merge them<br><br>Select by attribute JORDKVALIT = 1 |

| **OFFSHORE** |
|---|

| **LOW EXCLUSION** | |
|---|---|
| **VARIABLE** | **DATASET** |
| 3 km buffer around coast and islands | https://kartkatalog.geonorge.no/metadata/kartverket/n250-kartdata/442cae64-b447-478d-b384-545bc1d9ab48<br><br>Title: N250 Kartdata<br><br>Folder: Arealdekke<br><br>Select by attribute: OBJTYPE = Kystkontur |
| Marine Protected Areas (IUCN = 1a) | https://www.protectedplanet.net/c/monthly-updates/2019/july-2019-update-of-the-wdpa<br><br>Select by attribute: MARINE = 1 AND IUCN_CAT = Ia |
| Coral Reef | https://kartkatalog.geonorge.no/metadata/fiskeridirektoratet/korallrev-forbudsomrader/3be8f59c-cf30-47b5-ab5d-61abab25942b<br><br>Title: Korallrev – forbudsområder<br><br>Format: SOSI |
| Naturetypes: Slåttmark and Slåttmyr | https://kartkatalog.miljodirektoratet.no/Dataset/Details/10<br><br>Chose Norge uten Svalbard and selected viktige naturtyper<br><br>Select by attribute: NATURTYPE = Slåtte – og beitemyr and Slåttemark |
| Ship traffic buffer: 500m<br><br>(Other boat routes, car ferry, passanger ferry) | https://kartkatalog.geonorge.no/metadata/kartverket/n250-kartdata/442cae64-b447-478d-b384-545bc1d9ab48<br><br>Title: N250 Kartdata<br><br>Folder: Samferdsel<br><br>Files: Annen-Båtrute, Bilferjestrekning, Passasjerferjestrekning |
| Fields and pipeline | Download data from:<br><br>http://factpages.npd.no/ReportServer?/FactPages/geography/geography_all&rs:Command=Render&rc:Toolbar=false&rc:Parameters=f&IpAddress=1&CultureCode=nb-no<br><br>Data description (Attributter):<br><br>http://factpages.npd.no/factpages/Default.aspx?culture=nb-no&nav1=wellbore |
| **NEIGHBOURS** | |
| **HIGH EXCLUSION** | |

| VARIABLE | DATASET |
|---|---|
| Cultural heritage: Kulturmiljøer, sikringssoner, lokaliteter, brannsmitteområder | https://kartkatalog.geonorge.no/metadata/riksantikvaren/kulturminner-kulturmiljoer/17adbcac-bbb2-4efc-ab51-756573c8f178<br><br>Title: Kulturminner – Kulturmiljøer<br><br>Format: FGDB |
| Walking routes, ski runs, biking routes other-routes – 2 km buffer | |

| FAUNA | |
|---|---|
| **HIGH RESTRICTION** | |
| **VARIABLE** | **DATASET** |
| IBA – Important Bird Areas 3 km buffer | |
| All areas important for biodiversity | https://kartkatalog.miljodirektoratet.no/Dataset/Details/10<br><br>Chose Norge uten Svalbard and select viktige naturtyper. |
| All wild reindeer areas | https://kartkatalog.geonorge.no/metadata/miljodirektoratet/villreinomrader/fc59e9a4-59df-4eb3-978a-1c173b84bf4e<br><br>Title: Vilreinområder<br><br>Format: FGDB |
| Arter av veldig stor og stor forvaltningsinteresse (Species of very important and important management) | https://kartkatalog.geonorge.no/metadata/miljodirektoratet/a8456aed-441a-40c4-831f-46bcbe4e6ff1<br><br>Format: GDB<br><br>Select by Attribute: BM_TAKSON_BMFORVALTNINGSKATEGORI = 1 (very important) = 2 (important) |
| **NATURE** | |
| **HIGH EXCLUSION** | |
| **VARIABLE** | **DATASET** |
| Strict nature reserve –<br><br>IUCN = 1a, 1b II<br><br>5 km buffer<br><br>+ all IUCN categories | https://kartkatalog.geonorge.no/metadata/uuid/5857ec0a-8d2c-4cd8-baa2-0dc54ae213b4<br><br>Title: Naturvernområder<br><br>Format: SOSI |
| Kystlinghei<br><br>5 km buffer | https://kartkatalog.miljodirektoratet.no/Dataset/Details/10<br><br>Chose Norge uten Svalbard and selected viktige naturtyper.<br><br>Select by attribute: NATURYPE = Kystlinghei |

| **Sámi REINDEER HERDING** | |
|---|---|
| **HIGH EXCLUSION** | |
| **VARIABLE** | **DATASET** |
| All areas used by reindeer:<br><br>Migratory zones,<br><br>All seasons grazing land, Concession permit regions,<br><br>Expropriation areas,<br><br>Grazing zones, Enclosed grazing sites, gathering sites | https://kartkatalog.geonorge.no/metadata/uuid/f9c1e228-892f-4f1a-9e4e-b6d6149f373c<br><br>Title: Reindrift – Flyttlei<br><br>Format: SOSI<br><br>https://kartkatalog.geonorge.no/metadata/landbruksdirektoratet/reindrift-arstidsbeite-vinterbeite/63f655ef-f625-43cf-a512-bb8164bf53a4<br><br>Title: Reindrift – Årstidsbeite – Vinterbeite<br><br>Format: SOSI<br><br>https://kartkatalog.geonorge.no/metadata/landbruksdirektoratet/reindrift-arstidsbeite-hostbeite/6383f5a8-3a4d-48fc-8c67-f1eeec24fd8b<br><br>Title: Reindrift – Årstidsbeite – Høstbeite<br><br>Format: SOSI<br><br>https://kartkatalog.geonorge.no/metadata/landbruksdirektoratet/reindrift-arstidsbeite-hostvinterbeite/85a4c5e3-25ab-427c-b664-bbac2d0c9e79<br><br>Title: Reindrift – Årstidsbeite – Høstvinterbeite<br><br>Format: SOSI<br><br>https://kartkatalog.geonorge.no/metadata/landbruksdirektoratet/reindrift-arstidsbeite-varbeite/fa02a652-cd6d-4828-9fb5-7bd4515aa6d0<br><br>Title: Reindrift – Årstidsbeite – Vårbeite<br><br>Format: SOSI<br><br>https://kartkatalog.geonorge.no/metadata/landbruksdirektoratet/reindrift-arstidsbeite-sommerbeite/d5d1e2d4-7dc0-47ce-8776-ff64b07d788e<br><br>Title: Reindrift – Årstidsbeite – Sommerbeite<br><br>Format: SOSI<br><br>https://kartkatalog.geonorge.no/metadata/landbruksdirektoratet/reindrift-konsesjonsomrade/49efb2b2-93e3-4175-b10b-65b509d73c2a<br><br>Title: Reindrift- Konsesjonsområde |

| | Format: SOSI |
| --- | --- |
| | https://kartkatalog.geonorge.no/metadata/landbruksdirektoratet/reindrift-ekspropriasjonsomrade/1c64c5ff-0069-4f8e-9a2b-948c7ce3d527 |
| | Title: |
| | Reindrift- Ekspropriasjonsområde |
| | Format: SOSI |
| | https://kartkatalog.geonorge.no/metadata/landbruksdirektoratet/reindrift-reinbeiteomrade/d02dc4bd-77d5-4b3b-a316-5a488b6fe811 |
| | Title: Reindrift – Reinbeiteområde |
| | Format: SOSI |
| | https://kartkatalog.geonorge.no/metadata/landbruksdirektoratet/reindrift-beitehage/df2db95d-adbc-4807-bb46-00b729caed7c |
| | Title: Reindrift- Beitehage |
| | Format: SOSI |
| | https://kartkatalog.geonorge.no/metadata/landbruksdirektoratet/reindrift-oppsamlingsomrade/a02e84ec-322c-47a7-a626-ca02d57d1f7e |
| | Title: Reindrift- Oppsamlingsområde |
| | Format: SOSI |

| **PV** | |
| --- | --- |
| **HIGH EXCLUSION** | |
| **VARIABLE** | **DATASET** |

| Land used for agriculture | https://kartkatalog.geonorge.no/metadata/norsk-institutt-for-biookonomi/arealressurskart-ar50-jordbruk/76255ebe-2a0e-401e-87c8-7618dd196cf2 |
| --- | --- |
| | Title: Arealressurskart – AR50 – Jordbruk |
| | Format: SOSI, GML |
| | SOSI: Akershus, Aust Agder, Buskerd, Finnmark, Hordaland, MøreRomsdalen, Oslo, Rogaland, Romsa Troms, Sogn Fjordane, Telemark, Vest Agder, Vestfold, Østfold |
| | GML: Hedmark, Nordland, Oppland |
| | Select by attribute: Artype = 20 |
| All soil qualities | https://kartkatalog.geonorge.no/metadata/norsk-institutt-for-biookonomi/jordkvalitet/35c38144-c0a0-4ed9-a66f-21b80bc17fa7 |
| | Title: Jordkvalitet |
| | Format: SOSI |
| | Needed to download every region by itself and then merge them |
| | Select by attribute JORDKVALIT = 1 and 2 |

| **OFFSHORE** | |
| --- | --- |
| **HIGH EXCLUSION** | |
| **VARIABLE** | **DATASET** |

| 10 km buffer around coast and islands | https://kartkatalog.geonorge.no/metadata/kartverket/n250-kartdata/442cae64-b447-478d-b384-545bc1d9ab48 |
| --- | --- |
| | Title: N250 Kartdata |
| | Folder: Arealdekke |
| | Select by attribute: OBJTYPE = Kystkontur |
| All marine Protected Areas<br><br>IUCN = 1a, 1b, 2 with 5 km buffer | https://www.protectedplanet.net/c/monthly-updates/2019/july-2019-update-of-the-wdpa<br><br>Select by attribute: MARINE = 1 AND IUCN_CAT = Ia, Ib and II |
| Coral Reef | https://kartkatalog.geonorge.no/metadata/fiskeridirektoratet/korallrev-forbudsomrader/3be8f59c-cf30-47b5-ab5d-61abab25942b |
| | Title: Korallrev – forbudsområder |
| | Format: SOSI |
| Nature types: Slåttmark and Slåttmyr | https://kartkatalog.miljodirektoratet.no/Dataset/Details/10 |
| | Chose Norge uten Svalbard and selected viktige naturtyper |
| | Select by attribute: NATURTYPE = Slåtte – og beitemyr and Slåttemark |
| Ship traffic buffer: 500m<br><br>(Other boat routes, car ferry, passenger ferry) | https://kartkatalog.geonorge.no/metadata/kartverket/n250-kartdata/442cae64-b447-478d-b384-545bc1d9ab48 |
| | Title: N250 Kartdata |
| | Folder: Samferdsel |
| | Files: Annen-Båtrute, Bilferjestrekning, Passasjerferjestrekning |

| Fields and pipeline | Download data from: http://factpages.npd.no/ReportServer?/FactPages/geography/geography_all&rs:Command=Render&rc:Toolbar=false&rc:Parameters=f&IpAddress=1&CultureCode=nb-no

Data description (Attributter):

http://factpages.npd.no/factpages/Default.aspx?culture=nb-no&nav1=wellbore |
|---|---|
| Fishing Areas | https://kartkatalog.geonorge.no/metadata/uuid/c6082425-8133-4f4d-bc46-8960c78232ce |